\documentclass[preprint,aps,prd,superscriptaddress,showpacs,nofootinbib]{revtex4}%
\usepackage{graphicx}
\usepackage{amsmath}
\usepackage{amssymb}
\usepackage{bm}
\usepackage{epsfig}
\usepackage{mathrsfs}
\usepackage{hyperref}

\begin{document}
\preprint{}
\title{\mbox{}\\[10pt]
New approach to the resummation of logarithms\\
in Higgs-boson decays to a 
vector quarkonium plus a photon\\ 
}
\author{Geoffrey~T.~Bodwin}
\email[]{gtb@anl.gov}
\affiliation{High Energy Physics Division, Argonne National Laboratory,
Argonne, Illinois 60439, USA}
\author{Hee~Sok~Chung}
\email[]{chungh@anl.gov}
\affiliation{High Energy Physics Division, Argonne National Laboratory,
Argonne, Illinois 60439, USA}
\author{June-Haak~Ee}
\email[]{chodigi@gmail.com}
\affiliation{Department of Physics, Korea University, Seoul 02841, Korea}
\author{Jungil~Lee}
\email[]{jungil@korea.ac.kr}
\affiliation{Department of Physics, Korea University, Seoul 02841, Korea}
\date{\today}
\begin{abstract}
We present a calculation of the rates for Higgs-boson decays to a vector
heavy-quarkonium state plus a photon, where the heavy quarkonium states
are the $J/\psi$ and the $\Upsilon(nS)$ states, with $n=$ 1, 2, or 3.
The calculation is carried out in the light-cone formalism, combined
with nonrelativistic QCD factorization, and is accurate at leading order
in $m_Q^ 2/m_H^2$, where $m_Q$ is the heavy-quark mass and $m_H$ is the
Higgs-boson mass. The calculation contains corrections through
next-to-leading order in the strong-coupling constant $\alpha_s$ and the
square of the heavy-quark velocity $v$, and includes a resummation of
logarithms of $m_H^2/m_Q^2$ at next-to-leading logarithmic accuracy. We
have developed a new method, which makes use of Abel summation,
accelerated through the use of Pad\'e approximants, to deal with
divergences in the resummed expressions for the quarkonium light-cone
distribution amplitudes. This approach allows us to make definitive
calculations of the resummation effects. Contributions from the
order-$\alpha_s$ and order-$v^2$ corrections to the light-cone
distribution amplitudes that we obtain with this new method differ
substantially from the corresponding contributions that one obtains from
a model light-cone distribution amplitude [M.~K\"onig and M.~Neubert,
J.\ High\ Energy\ Phys.\ 08\ (2015)\ 012]. Our results for the real parts of the
direct-process amplitudes are considerably smaller than those from one
earlier calculation   [G.\,T.~Bodwin, H.\,S.~Chung, J.-H.~Ee, J.~Lee,
and F.~Petriello, Phys.\ Rev.\ D {\bf 90}, 113010 (2014)], reducing the
sensitivity to the Higgs-boson--heavy-quark couplings, and  are somewhat
smaller than those from another earlier calculation [M.~K\"onig and
M.~Neubert, J.\ High\ Energy\ Phys.\ 08\ (2015)\ 012]. However, our results for the
standard-model Higgs-boson branching fractions are in good agreement
with those in M.~K\"onig and M.~Neubert, 
J.\ High\ Energy\ Phys.\ 08\ (2015)\ 012.

\end{abstract}
\maketitle

\section{Introduction}

Several years ago, it was pointed out that Higgs-boson ($H$) decays into
a vector charmonium state ($V$) plus a photon ($\gamma$) proceed through
two processes \cite{Bodwin:2013gca}. One process is the ``direct
process,'' in which the Higgs boson decays into a heavy quark-antiquark
($Q\bar Q$) pair, followed by the radiation of a real photon by the $Q$
or $\bar Q$ and the subsequent evolution of the $Q\bar Q$ pair into the
quarkonium. The other process is the ``indirect process,'' in which
the Higgs boson decays via a $W$-boson loop or a quark loop into a
$\gamma$ and a virtual photon ($\gamma^*$), followed by the decay of the
$\gamma^*$ into a $Q\bar Q$ pair, which evolves into the quarkonium.

The direct amplitude is proportional to the $HQ\bar Q$ coupling.
However, its standard-model (SM) value is generally too small to lead to
a rate that is measurable at the LHC. In the case in which the
quarkonium is a $J/\psi$, the SM indirect amplitude is much larger than
the SM direct amplitude and leads to a rate that is potentially
measurable in a high-luminosity LHC \cite{Bodwin:2013gca}. Furthermore,
the contribution from interference between the direct and indirect
amplitudes, which is destructive, may also be within the realm of
measurement at a high-luminosity LHC \cite{Bodwin:2013gca} and could
lead to a determination of the $Hc\bar c$ coupling. In the cases in
which the quarkonium is an $\Upsilon(nS)$ state, the SM rates are too
small to be measured even at a high-luminosity LHC
\cite{Bodwin:2013gca}. However, owing to the destructive interference
between the direct and indirect amplitudes, the rates are very sensitive
to deviations of the direct amplitudes from the SM values
\cite{Bodwin:2013gca}. Because the direct and indirect amplitudes for
the decays $H\to V+\gamma$ are comparable in size, these decays can give
information about the phases of the $HQ\bar Q$ couplings. They are the
only processes that have been identified so far that can yield that phase
information.

The indirect amplitude can be obtained, up to corrections of relative 
order $m_Q^2/m^{2}_H$, from the amplitude for $H\to 
\gamma\gamma$ \cite{Bodwin:2013gca}, which is known in the SM with a 
precision of a few percent \cite{Dittmaier:2011ti,Dittmaier:2012vm}. 
Here, $m_Q$ is the heavy-quark mass and $m_H$ is the Higgs-boson mass.

In Ref.~\cite{Bodwin:2013gca}, the direct amplitude was computed through
next-to-leading order (NLO) in the strong coupling $\alpha_s$ by making
use of the result of Shifman and Vysotsky \cite{Shifman:1980dk}.
That result was derived by making use of light-cone methods
\cite{Lepage:1980fj,Chernyak:1983ej} that are valid up to corrections of
order $m_Q^2/m_H^2$. In addition, in Ref.~\cite{Bodwin:2013gca},
logarithms of $m_H^2/m_Q^2$ were resummed at leading logarithmic (LL)
accuracy to all orders in $\alpha_s$ by making use of the LL resummed
expression for the direct amplitude in Ref.~\cite{Shifman:1980dk}.

The largest single uncertainty in the calculation of
Ref.~\cite{Bodwin:2013gca} was due to uncalculated relativistic
corrections to the direct amplitude of relative order $v^2$, where $v$
is the velocity of the $Q$ or $\bar Q$ in the quarkonium rest frame.
Those order-$v^2$ corrections were computed in Ref.~\cite{Bodwin:2014bpa}
in the nonrelativistic QCD (NRQCD) formalism \cite{Bodwin:1994jh} and,
also, in the light-cone formalism \cite{Lepage:1980fj,Chernyak:1983ej},
so as to make contact with the light-cone calculation of
Ref.~\cite{Shifman:1980dk}. 

Logarithms of $m_H^2/m_Q^2$ can be resummed by evolving the $HQ\bar Q$
coupling, which is proportional to $m_Q(\mu)$, the quarkonium decay
constant, and the light-cone distribution amplitude (LCDA) from the
renormalization scale $\mu=m_Q$ to the renormalization scale $\mu=m_H$.
The standard method for carrying out the evolution of the LCDA is to
expand the LCDA in a series of eigenfunctions of the lowest-order
evolution kernel. The eigenfunctions are proportional to Gegenbauer
polynomials \cite{Brodsky:1980ny}. In Ref.~\cite{Bodwin:2014bpa}, it was
noticed that the eigenfunction series is not convergent in the case of
the order-$v^2$ corrections to the direct amplitude. Consequently, for
the order-$v^2$ correction, logarithms of $m_H^2/m_Q^2$ were summed only
through relative order $\alpha_s^2$ in Ref.~\cite{Bodwin:2014bpa}.

Resummation of logarithms of $m_H^2/m_Q^2$ at
next-to-leading-logarithmic (NLL) accuracy requires a calculation in the
light-cone formalism of the order-$\alpha_s$ corrections to both the
hard-scattering kernel for the direct process and the LCDA. That calculation
was accomplished in Ref.~\cite{Wang:2013ywc} at leading order (LO) in
$v$. (The calculation of the order-$\alpha_s$ correction to the 
hard-scattering kernel in Ref.~\cite{Wang:2013ywc} was
confirmed in Ref.~\cite{Koenig:2015pha}.) The calculation of the LCDA
was carried out in the NRQCD framework, and the result was expressed in
terms of the NRQCD nonperturbative long-distance matrix elements (LDMEs)
\cite{Jia:2008ep}.

The actual resummation of logarithms of $m_H^2/m_Q^2$ at NLL
accuracy was carried out in Ref.~\cite{Koenig:2015pha}, in which it was
found that the NLL corrections have a substantial impact on the
numerical results for the rates. In that work, the calculational
strategy involved introducing a model LCDA whose nonzero second moment
would take into account the known order-$v^2$ and order-$\alpha_s$
corrections to the LCDA at a scale of $1$~GeV. This approach avoids
the problem of the lack of convergence of the eigenfunction expansion in
a calculation of the order-$v^2$ corrections to the LCDA. However, as we
will see, the model wave function does not give a very accurate
accounting of the order-$v^2$ and order-$\alpha_s$ corrections to the
LCDA, even after evolution to the scale $m_H$.

In this paper, we present a new method for calculating the evolution of
the order-$v^2$ corrections to the LCDA. The method introduces a
regulator that defines the generalized functions (distributions) that
appear in the initial LCDAs as sequences of ordinary functions. The
regulator method is equivalent to Abel summation of the eigenfunction
expansion. In order to accelerate the convergence of the Abel
summation, we introduce Pad\'e approximants to obtain an approximate
analytic continuation in the regulator variable that converges rapidly
as the regulator is removed. We refer to this method that makes use of a
combination of Abel summation and Pad\'e approximants as the
``Abel-Pad\'e method.'' The Abel-Pad\'e method gives very
accurate results in cases for which analytic results are known for the
LCDAs, even in situations in which the eigenfunction expansion
diverges. The Abel-Pad\'e method solves the general problem of
carrying out the scale evolution in a nonrelativistic expansion of the
LCDA for heavy-quarkonium systems, and it should be applicable in other
situations in which series of orthogonal polynomials fail to converge.

The results that we obtain with the Abel-Pad\'e method agree
reasonably well with the perturbative estimates of
Ref.~\cite{Bodwin:2014bpa}. However, the Abel-Pad\'e method gives
results that differ significantly from those that are obtained by making
use of the model of Ref.~\cite{Koenig:2015pha}. We use the
Abel-Pad\'e method to obtain a complete calculation of the
rates for $H\to V+\gamma$ through orders $\alpha_s$ and $v^2$ and to all
orders in $\alpha_s$ through order $v^2$ at NLL accuracy.

The remainder of this paper is organized as follows. In
Sec.~\ref{sec:light-cone}, we discuss the light-cone amplitude for the
direct process through orders $\alpha_s$ and $v^2$. In
Sec.~\ref{sec:resummation}, we describe the resummation of logarithms of
$m_H^2/m_Q^2$ and give resummed expressions for the contributions
to the direct amplitude in terms of sums over eigenfunctions of the LO
evolution kernel. Section~\ref{sec:pade} contains a discussion of the 
problem of the nonconvergence of the eigenfunction series and a 
presentation of a solution of the problem, which leads to the 
Abel-Pad\'e method for summing the series. In
Sec.~\ref{sec:model}, we compare results from the Abel-Pad\'e
method with those that follow from the model LCDA that was proposed in
Ref.~\cite{Koenig:2015pha}. In Sec.~\ref{sec:computation} we give the
expressions that we use to compute the direct amplitudes and the
indirect amplitudes and discuss the numerical inputs that we use and the
sources of uncertainties. We also present a novel method to compute
uncertainties in the decay rates that allows us to deal with the highly
nonlinear dependences of the decay rates on the input parameters. We
give our numerical results in Sec.~\ref{sec:results}, and we summarize
and discuss our results in Sec.~\ref{sec:summary}.

\section{Light-cone amplitude for the direct process 
\label{sec:light-cone}}

In the light-cone approach, the direct amplitude for $H\to V+\gamma$ is 
given, up to corrections of relative order $m_Q^2/m_{H}^2$,
by\footnote{See, for example, Ref.~\cite{Bodwin:2013gca}.}
\begin{eqnarray}
i {\cal M}_{\rm dir}^{\rm LC} [H\to V+\gamma]
&=& \frac{i}{2} e e_Q \kappa_Q \overline{m}_Q (\mu) 
( \sqrt{2} G_F)^{1/2} f_V^\perp (\mu)
\left( - \epsilon^*_V \cdot \epsilon^*_\gamma 
+ \frac{\epsilon^*_V \cdot p_\gamma p \cdot \epsilon_\gamma^*}
{p_\gamma \cdot p} \right) 
\nonumber \\ && \times 
\int_0^1 dx\, T_H (x,\mu) \phi_V^\perp(x,\mu),
\label{M-direct}
\end{eqnarray}
where $e$ is the electric charge, $e_Q$ is the fractional charge of the
heavy quark $Q$, $\kappa_Q$ is an adjustable parameter in the
$HQ\bar{Q}$ coupling  whose SM value is 1,  $\overline{m}_Q$ is the mass
of $Q$ in the modified minimal subtraction
($\overline{\textrm{MS}}$) scheme, $G_F$ is the Fermi constant,
$f_V^\perp$ is the decay constant of the vector quarkonium $V$,
$\epsilon_V$ and $p$ are the quarkonium polarization and momentum,
respectively, $\epsilon_\gamma$ and $p_\gamma$ are the photon
polarization and momentum, respectively, $\mu$ is the renormalization
scale, and $x$ is the $Q\bar Q$ momentum fraction of $V$, which runs
from $0$ to $1$. $\phi_V^\perp(x,\mu)$ is the vector-quarkonium LCDA,
which is defined by
\begin{equation}
\frac{1}{2} \langle V | \bar Q(z) [\gamma^\mu, \gamma^\nu] [z,0] Q(0)
|0\rangle
= 
f_V^\perp(\mu)
(\epsilon_V^{*\mu} p_V^\nu - \epsilon_V^{*\nu} p_V^\mu ) 
\int_0^1 dx\, e^{i p^- z x} 
\phi_V^\perp(x,\mu)
\end{equation}
and has the normalization $\int_0^1 dx\, \phi_V^\perp(x, \mu) = 1$.
The coordinate $z$ lies along the plus light-cone direction, and
the gauge link
\begin{equation}
[z,0]=P\,\exp\left[ig_s\int_0^z dx \,A^+_a(x)T^a\right]
\label{gauge-link}
\end{equation}
makes the nonlocal operator gauge invariant. In
Eq.~(\ref{gauge-link}), $g_s=\sqrt{4\pi\alpha_s}$, $A^\mu_a$ is the
gluon field with the color index $a=1,$ 2, $...$, $N_c^2-1$, $T^a$
is the generator of the fundamental representation of $\textrm{SU}(N_c)$
color, and the symbol $P$ denotes path ordering. The
nonrelativistic expansion of $\phi_V^\perp(x, \mu)$, through linear
orders in $\alpha_s$ and $v^2$, is
\begin{equation}
\label{eq:lcda}%
\phi_V^\perp(x,\mu) = 
\phi_V^{\perp (0)}(x,\mu)
+ 
\langle v^2 \rangle_V 
\phi_V^{\perp (v^2)}(x,\mu)
+
\frac{\alpha_s(\mu)}{4 \pi} 
\phi_V^{\perp (1)}(x,\mu) +O(\alpha_s^2, \alpha_s v^2, v^4), 
\end{equation}
where the LO contribution is given by
\begin{equation}
\phi_V^{\perp (0)}(x,\mu)=\delta(x-\tfrac{1}{2})
\label{phi-V-perp-0}
\end{equation}
and $\delta$ is the Dirac delta function. $\langle v^2\rangle_V$ is
proportional to the ratio of the NRQCD LDME of order $v^2$ to the
LDME of order $v^0$:
\begin{equation}
\langle v^2\rangle_V
=\frac{1}{m_Q^2}
\frac{
\langle V(\bm{\epsilon}_V)|
\psi^\dagger (-\tfrac{i}{2}\stackrel{\leftrightarrow}{\bm{\nabla}}
)^2 \bm{\sigma}\cdot\bm{\epsilon}_V\chi|0\rangle
}
{
\langle V(\bm{\epsilon}_{V})|
\psi^\dagger \bm{\sigma}\cdot\bm{\epsilon}_V\chi|0\rangle
}.
\label{v2ME}
\end{equation}
Here, $\psi$ is the two-component (Pauli) spinor field that annihilates
a heavy quark, $\chi^\dagger$ is the two-component spinor field that
annihilates a heavy antiquark, $\sigma_i$ is a Pauli matrix,
$|V(\bm{\epsilon}_V)\rangle$ denotes the vector quarkonium state in the
quarkonium rest frame with spatial polarization $\bm{\epsilon}_V$, and
$m_Q$ denotes the quark pole mass. The coefficient of the order-$v^2$
contribution, $\phi_V^{\perp (v^2)}$, was computed in
Ref.~\cite{Bodwin:2014bpa} and is given by
\begin{equation}
\phi_V^{\perp (v^2)}(x,\mu)=\frac{1}{24} \delta^{(2)} (x-\tfrac{1}{2}),
\label{phi-V-perp-vsq}
\end{equation}
where $\delta^{(n)}$ is the $n$th derivative of the Dirac delta function.
The coefficient of the order-$\alpha_s$ contribution, $\phi_V^{\perp
(1)}(x,\mu)$, was computed in Ref.~\cite{Wang:2013ywc} and is
given by\footnote{Equation~(3.17) of Ref.~\cite{Wang:2013ywc} applies to the
case in which $\Delta$ in Eq.~(3.16) of Ref.~\cite{Wang:2013ywc} is set
equal to zero. We thank the authors of Ref.~\cite{Wang:2013ywc} for
confirming that this is the case.}
\begin{eqnarray}  
\phi_V^{\perp (1)} (x,\mu)&=& 
C_F\theta(1-2x)
\bigg\{ \bigg[ \frac{8 x}{1-2 x}\bigg( \log \frac{\mu^2}{m_{Q}^2 (1-2 x)^2}-1 
\bigg) \bigg]_+
+ \bigg[ \frac{16 x (1- x) }{(1-2 x)^2} \bigg]_{++} 
\bigg\}\nonumber\\
&&+(x\leftrightarrow 1-x), 
\label{phi-V-perp-1}
\end{eqnarray}
where $C_F=(N_c^2-1)/(2N_c)$, $N_c=3$ is the number of colors, and the
plus and plus-plus distributions are defined by
\begin{subequations}
\begin{eqnarray}
\int_0^1 dx\, f(x)[g(x)]_{+\phantom{+}}  &=& 
\int_0^1 dx \, [f(x)-f(\tfrac{1}{2})]g(x), 
\\ 
\int_0^1 dx\, f(x)[g(x)]_{++}  &=& 
\int_0^1 dx \, [f(x)-f(\tfrac{1}{2})-f'(\tfrac{1}{2}) (x-\tfrac{1}{2})]g(x) . 
\end{eqnarray}
\end{subequations}
Although $\phi_V^{\perp (0)}(x,\mu)$ and  $\phi_V^{\perp 
(v^2)}(x,\mu)$ are independent of $\mu$, we keep $\mu$ explicit in their 
arguments as a reminder that a single scale $\mu$ applies to all of the terms 
in $\phi_V^{\perp}(x,\mu)$ [Eq.~(\ref{eq:lcda})].

The quarkonium decay constant $f_V^\perp(\mu)$ is given by 
\begin{equation}
f_V^\perp(\mu)= 
\frac{\sqrt{2 N_c} \sqrt{2 m_V}}{2 m_Q} \Psi_V(0)
\bigg[ 1 - \frac{5}{6} \langle v^2 \rangle_V 
- \frac{C_F \alpha_s(\mu)}{4 \pi} 
\bigg(\log \frac{\mu^2}{m_Q^2}+8 \bigg) 
+ O(\alpha_s^2, \alpha_s v^2, v^4) 
\bigg],
\label{f-V-perp}
\end{equation}
where the order-$v^2$ term was computed in Ref.~\cite{Bodwin:2014bpa} and 
the order-$\alpha_s$ term was computed in Ref.~\cite{Wang:2013ywc}.
Here, $\Psi_V(0)$ is the quarkonium wave function at the origin, which is 
given in terms of an NRQCD LDME by \cite{Bodwin:1994jh}
\begin{equation}                                          
\Psi_V(0)=\frac{1}{\sqrt{2N_c}}                                 
\langle V(\bm{\epsilon}_V)                                      
|\psi^\dagger \bm{\sigma}\cdot\bm{\epsilon}_V \chi|0\rangle.
\end{equation}

The hard-scattering kernel $T_H$ for the process $H \to V+\gamma$ is given by
\begin{subequations}
\label{T-H}%
\begin{equation}
T_H(x,\mu)=T_H^{(0)}(x,\mu)
+\frac{\alpha_s(\mu)}{4\pi}T_H^{(1)}(x,\mu)+
{\cal O}(\alpha_s^2),
\end{equation}
where
\begin{eqnarray}
T_H^{(0)}(x,\mu)&=&\frac{1}{x (1-x)},\\ 
T_H^{(1)}(x,\mu)&=&C_F\frac{1}{x (1-x)} 
\bigg[ 2 \bigg( \log \frac{m_{H}^2}{\mu^2} 
-i \pi \bigg) \log x (1-x) 
+ \log^2 x+\log^2 (1-x) -3 
\bigg].\nonumber\label{T-H-1}\\
\end{eqnarray}
\end{subequations}
The order-$\alpha_s$ term in $T_H$ was computed in
Ref.~\cite{Wang:2013ywc} by taking the quark mass to be the pole mass 
and in Ref.~\cite{Koenig:2015pha} by taking the quark mass to be the 
$\overline{\rm MS}$ mass.\footnote{Equation~(4.23) of
Ref.~\cite{Wang:2013ywc} contains a typo: $3\ln[\mu^2/(-m_h^2)]$ should
be replaced with $3\ln(\mu^2/m_Q^2)$. This typo was noted in
Ref.~\cite{Koenig:2015pha}. We thank the authors of
Ref.~\cite{Wang:2013ywc} for confirming the existence of this typo.} The
expression in Eq.~(\ref{T-H-1}) is for the case in which the quark mass is 
taken to be the $\overline{\rm MS}$ mass.

\section{Resummation of logarithms in the direct amplitude 
\label{sec:resummation}}

Our strategy for resumming logarithms of $m_H^2/m_Q^2$ is the following.
In Eq.~(\ref{M-direct}) we take the scale $\mu$ to be $m_H$. Then
$T_H$ [Eq.~(\ref{T-H})] contains no large logarithms. Note that, if one
takes the quark mass in the computation of $T_H$ to be the pole mass,
then the order-$\alpha_s$ correction to $T_H(x,\mu)$ contains a term
that is proportional to $\log(m_H^2/m_Q^2)$, as can be seen from the
corrected version of Eq.~(4.23) of Ref.~\cite{Wang:2013ywc}. Such large
NLLs would slow, or even spoil, the convergence of the perturbation
expansion. We initially evaluate $\phi_V^\perp(x,\mu)$ and
$f_V^\perp(\mu)$ at a scale $\mu_0$ of order $m_Q$, so that the
perturbative expressions in Eqs.~(\ref{phi-V-perp-0}),
(\ref{phi-V-perp-vsq}) and (\ref{phi-V-perp-1}) do not contain any
logarithms of $m_H^2/m_Q^2$. Then, we evolve $\phi_V^\perp(x,\mu)$ and
$f_V^\perp(\mu)$ to the scale $\mu=m_H$, along with $\overline{m}_Q
(\mu)$. Expressions for the evolution of $\overline{m}_Q (\mu)$ and
$f_V^\perp(\mu)$ are given in Appendix~\ref{sec:mass-decayconst-evolve}.
We now address the evolution of $\phi_V^\perp(x,\mu)$.

\subsection{Evolution of the LCDA}

The LCDA $\phi_V^\perp(x,\mu)$ satisfies the evolution equation 
\cite{Lepage:1980fj} 
\begin{equation}
\mu^2 \frac{\partial}{\partial \mu^2}
\phi_V^\perp(x,\mu) =  C_F
\frac{\alpha_s(\mu) }{2\pi}
\int_{0}^1\! dy\, V_T (x,y) 
\phi_V^\perp(y,\mu),
\label{phivperp-evolution}
\end{equation}
where the LO evolution kernel $V_T(x,y)$ is given by
\cite{Lepage:1980fj}
\begin{subequations}
\begin{eqnarray}
V_T(x,y) &=& V_0 (x,y) 
- \frac{1-x}{1-y} \theta (x-y)
- \frac{x}{y} \theta(y-x),
\\
V_0(x,y) &=& V_{\rm BL} (x,y) 
- \delta(x-y) \int_{0}^1\! dz\, V_{\rm BL} (z,x),
\\
V_{\rm BL} (x,y)
&=& \frac{1-x}{1-y} \left( 1+ \frac{1}{x-y} \right) \theta (x-y)
+ \frac{x}{y} \left( 1+ \frac{1}{y-x} \right) \theta(y-x).
\end{eqnarray}
\end{subequations}

As is well known, the eigenfunctions of LO evolution kernel
for $\phi_V^\perp(x,\mu)$ are given by \cite{Brodsky:1980ny}
\begin{equation}
G_n(x) = w(x) C_n^{(3/2)} (2 x-1), 
\label{eigenfunctions}
\end{equation}
where $w(x) = x (1-x)$ is the weighting function and the $C_n^{(3/2)}$
are Gegenbauer polynomials. The corresponding eigenvalues (anomalous
dimensions) are
\begin{equation}
\gamma_n^{\perp(0)}=8C_F(H_{n+1}-1), 
\end{equation}
where the $H_n$ are harmonic numbers.
The orthogonality relation of the Gegenbauer polynomials is given by
\begin{eqnarray}
N_n \int_0^1 dx \, w(x)\,C_n^{(3/2)}(2x-1) 
C_m^{(3/2)} (2 x-1) 
&=& N_n \int_0^1 dx \, G_n(x)\, C_m^{(3/2)} (2 x-1)
\nonumber \\
&=& \delta_{nm},
\label{orthonormal}
\end{eqnarray}
where the normalization factor $N_n$ is given by
\begin{equation}
N_n = \frac{4 (2 n+3)}{(n+1) (n+2)}.
\end{equation}

In order to work out the evolution of the LCDAs, it is convenient to 
write them in terms of the eigenfunctions. Using 
Eq.~(\ref{orthonormal}), we have 
\begin{subequations}
\begin{equation}
\phi_V^\perp (x,\mu) = \sum_{n=0}^\infty \phi_{n}^\perp (\mu) G_n(x), 
\end{equation}
where the moments $\phi_n^\perp (\mu)$ are given by
\begin{equation}
\phi_{n}^\perp (\mu)
= 
N_n 
\int_0^1 dx \, C_n^{(3/2)}(2 x-1)\phi_V^\perp(x,\mu). 
\end{equation}
\end{subequations}
In a similar fashion, we can write $T_H$ in terms of Gegenbauer
polynomials:
\begin{subequations}
\begin{equation}
T_H (x,\mu) = \sum_{n=0}^\infty N_n T_n(\mu) C_n^{(3/2)}(2x-1), 
\end{equation}
where 
\begin{equation}
T_n(\mu)
= \int_0^1 dx \, T_H(x,\mu)G_n(x). 
\end{equation}
\end{subequations}
Then, using Eq.~(\ref{orthonormal}), we can write the light-cone
amplitude, at least formally, as a sum over moments of $T_H$ and
$\phi_V^\perp$:
\begin{equation}
\int_0^1 dx\, T_H(x,\mu)\phi_V^\perp(x,\mu)=\sum_{n=0}^\infty 
T_n(\mu)\phi_n^\perp(\mu).
\label{moment-sum}
\end{equation}

The moments $\phi_n^\perp(\mu)$ can be written in terms of the moments 
$\phi_n^\perp(\mu_0)$ as
\begin{equation}
\phi_n^\perp(\mu)=\sum_{k=0}^n U_{nk}(\mu,\mu_0)\phi_{k}^\perp(\mu_0),
\label{moment-evolution}
\end{equation}
where we are using the notation of Ref.~\cite{Koenig:2015pha}. The
expressions for $ U_{nk}(\mu,\mu_0)$ at LL and  NLL accuracies
are given in Appendix~\ref{sec:evolution-matrix}. Note that the
off-diagonal elements of $U_{nk}(\mu,\mu_0)$ are nonvanishing only for
even $n-k$~\cite{Mueller:1993hg, Mueller:1994cn}.

We decompose the light-cone amplitude according to the powers of 
$\alpha_s$ and $v^2$:
\begin{subequations}
\begin{eqnarray}
\int_0^1 dx \, T_H(x,\mu) \phi_V^\perp(x,\mu)
&=& 
{\cal M}^{(0,0)}(\mu) + 
\frac{\alpha_s (\mu) }{4 \pi} {\cal M}^{(1,0)}(\mu) + 
\frac{\alpha_s (m_Q) }{4 \pi}{\cal M}^{(0,1)}(\mu)
\nonumber \\
&& + \langle v^2 \rangle_V {\cal M}^{(0,v^2)}(\mu)
+ O(\alpha_s^2, \alpha_s v^2, v^4),
\end{eqnarray}
where
\begin{equation}
{\cal M}^{(i,j)}(\mu)=\int_0^1 dx\, T_H^{(i)}(x,\mu)\phi_V^{\perp(j)}(x,\mu)=
\sum_{n=0}^\infty T_n^{(i)}(\mu)\phi_n^{\perp(j)}(\mu).
\label{M-eigenvalue-series}
\end{equation}
\end{subequations}
$T_n^{(0)}(\mu)$ and $T_n^{(1)}(\mu)$ vanish for $n$ odd and are given 
for $n$ even by
\begin{subequations}
\begin{eqnarray}
T_n^{(0)}(\mu)&=&1,
\\
T_n^{(1)}(\mu)/C_F&=&
-4 (H_{n+1}-1) \bigg( \log \frac{m_{H}^2}{\mu^2} -i \pi \bigg) + 
4H_{n+1}^2-3+4 \pi i,
\end{eqnarray}
\end{subequations}
where the expression for $T_n^{(1)}(\mu)$ was first given in
Ref.~\cite{Koenig:2015pha}. The $\phi_n^{\perp(i)}(\mu)$ also vanish for $n$
odd.

For ${\cal M}^{(0,0)}(\mu)$, we use the NLL expression for
$U_{nk}(\mu,\mu_0)$ to compute $\phi_n^{\perp(0)}(\mu)$, while, for the
other ${\cal M}^{(i,j)}(\mu)$, we use the LL expression for
$U_{nk}(\mu,\mu_0)$.

As was noted in the appendix of Ref.~\cite{Bodwin:2014bpa}, the 
eigenfunction series for ${\cal M}^{(0,v^2)}(\mu)$ is not convergent. Some 
of the eigenfunction series for the other ${\cal M}^{(i,j)}(\mu)$ 
converge rather slowly. We address these issues of nonconvergence and 
slow convergence in Sec.~\ref{sec:pade}.

\section{Nonconvergence of the eigenfunction series and summation by the
Abel-Pad\'e method \label{sec:pade}}

\subsection{The problem of nonconvergence}

From the theory of orthogonal polynomials on a finite interval, we know
that a series of Gegenbauer polynomials $C_n^{(3/2)}(2x-1)$ can
represent sufficiently smooth functions over the interval $0 < x < 1$.
That is, $C_n^{(3/2)}(2x-1)$ are a complete set of functions and satisfy
the completeness relation
\begin{equation}
\sum_{n=0}^\infty 
N_n w(x) C_n^{(3/2)} (2 x-1)
 C_n^{(3/2)} (2 y-1) 
= \delta(x-y).
\label{completeness}
\end{equation}
It follows that the sum
over $n$ on the right side of Eqs.~(\ref{moment-sum}) or
(\ref{M-eigenvalue-series}) is well defined and is equal to the left
side of Eqs.~(\ref{moment-sum}) or (\ref{M-eigenvalue-series}) when
$T_H(x,\mu)$ and $\phi_V^\perp(x,\mu)$ are sufficiently smooth
functions of $x$ \cite{szego}. A difficulty can arise because the
nonrelativistic expansion of $\phi_V^\perp(x,\mu)$ contains generalized
functions (distributions) in $x$ about the point $x=1/2$. For example,
the factor $\delta^{(2)}(x-\tfrac{1}{2})$ in $\phi_V^{\perp (v^2)}$
[Eq.~(\ref{phi-V-perp-vsq})] causes the sum over $n$ in the expression
for ${\cal M}^{(0,v^2)}(\mu)$ to diverge, as was shown in the appendix
of Ref.~\cite{Bodwin:2014bpa}. Nevertheless, ${\cal M}^{(0,v^n)}(\mu)$
remains well defined as $\mu$ evolves.

In order to demonstrate this, we define the quantity
\begin{equation}
{\cal M}^{(i,j)}(\mu_f,\mu)=\int_0^1 dx\, T_H^{(i)}(x,\mu_f)
\phi_V^{\perp(j)}(x,\mu),
\label{Mmufmu-def}
\end{equation}
which gives the projection of $\phi_V^{\perp(j)}(x,\mu)$ onto the 
hard-scattering amplitude evaluated at the final scale in the evolution 
$\mu_f$. Note that ${\cal M}^{(i,j)}(\mu_f,\mu_f)={\cal M}^{(i,j)}(\mu_f)$. 
Now, ${\cal M}^{(0,v^n)}(\mu_f,\mu)$ satisfies the same evolution equation
as does $\phi_V^\perp(x,\mu)$, namely,
\begin{equation}
\mu^2 \frac{\partial}{\partial \mu^2}
{\cal M}^{(0,v^n)}(\mu_f,\mu) =  C_F
\frac{\alpha_s(\mu) }{2\pi}
\int_0^1 dx\, \int_{0}^1\! dy\, T_H^{(0)}(x,\mu_f) V_T (x,y) 
\phi_V^{\perp (v^n)}(y,\mu).
\label{Mmufmu-evolution}
\end{equation}
First, we note that ${\cal M}^{(0,v^n)}(\mu_f,\mu_0)$ is well defined.
This follows from the definition of ${\cal
M}^{(0,v^n)}(\mu_f,\mu_0)$ in Eq.~(\ref{Mmufmu-def}), the fact that
$\phi_V^{\perp (v^n)}(x,\mu_0)$ is proportional to
$\delta^{(n)}(x-\tfrac{1}{2})$, and the fact that $T_H^{(0)}(x,\mu_f)$ is
infinitely differentiable at $x=1/2$. [We remind the reader that
$T_H^{(0)}(x,\mu)$ is actually independent of $\mu$.] Furthermore, it is
easy to see that $\int_0^1 dx\, T_H^{(0)}(x,\mu)V_T(x,y)$ is infinitely
differentiable with respect to $y$ at $y=1/2$. It then follows from the
evolution equation (\ref{Mmufmu-evolution}) that $\mu^2
(\partial/\partial \mu^2) {\cal M}^{(0,v^n)}(\mu_f,\mu)$ is well defined
for all $\mu$ between $\mu_0$ and $\mu_f$. Therefore,  ${\cal
M}^{(0,v^n)}(\mu_f,\mu_f)= {\cal M}^{(0,v^n)}(\mu_f)$ is well defined.

\subsection{Solution of the problem and the Abel-Pad\'e method}

In order to address the difficulty of nonconvergent eigenfunction
series, we first define a smearing function $S(x,y,z)$ by modifying
the completeness relation (\ref{completeness}). We introduce a factor
$z^n$ into each term in the sum over $n$:
\begin{equation}
S(x,y,z) = \sum_{n=0}^\infty z^n
N_n w(x) C_n^{(3/2)} (2 x-1) 
C_n^{(3/2)} (2 y-1),
\label{smearing}
\end{equation}
where $z$ is a complex parameter. 
For $|z|<1$, the sum over $n$ in Eq.~(\ref{smearing}) is 
absolutely convergent, and $S(x,y,z)$ is an ordinary function of $x$ and 
$y$. As $z$ approaches $1$, $S(x,y,z)$ becomes more and more sharply peaked 
around $x=y$ and, in the limit $z\to 1$, is a representation of
$\delta(x-y)$.
We use the smearing function to define a smeared distribution amplitude:
\begin{eqnarray}
\phi_S(x,z,\mu) &=& 
\int_0^1 dy\,S(x,y,z) \phi_V^\perp (y,\mu) 
\nonumber \\
&=& 
\sum_{n=0}^\infty \phi_n^\perp(\mu)
\sum_{m=0}^{\infty} z^m w(x) C_m^{(3/2)}(2 x-1)
N_m \int_0^1 dy 
\,w(y) C_m^{(3/2)}(2 y-1) C_n^{(3/2)} (2 y-1) 
\nonumber \\
&=& 
\sum_{n=0}^\infty \phi_n^\perp(\mu)
\sum_{m=0}^{\infty} z^m w(x) C_m^{(3/2)}(2 x-1)
\delta_{nm} 
\nonumber \\
&=&\sum_{n=0}^\infty \phi_n^\perp(\mu) z^n G_n(x),
\label{smeared-dist}
\end{eqnarray}
where we have used the orthogonality relation (\ref{orthonormal}). For
$|z|<1$, $\phi_S(x,z,\mu)$ is an ordinary function of $x$. Because
$S(x,y,z)$ is a representation of $\delta(x-y)$ in the limit $z\to 1$,
$\phi_S(x,z,\mu)$ is a representation of $\phi_V^\perp(x,\mu)$ in the
limit $z\to 1$. That is, Eq.~(\ref{smeared-dist}) can be used to
define generalized functions in $\phi_V^\perp(x,\mu)$ as a limit of a
sequence of ordinary functions. It then follows, from the theory of
orthogonal functions, that, for any $z<1$,
\begin{equation}
\int_0^1 dx\, T_H(x,\mu)\phi_S(x,z,\mu)=\sum_{n=0}^\infty 
T_n(\mu) z^n \phi_n^\perp(\mu).\footnote{It can be seen from the analysis of the appendix of
Ref.~\cite{Bodwin:2014bpa} that, for $\phi_V^\perp(x,\mu) \to
\phi_V^{\perp (0)}(x,\mu)\equiv\delta^{(0)}(x-\tfrac{1}{2})$ and
$T_H(x,\mu)\to T_H^{(0)}(x,\mu)$, the sum on the right side of
Eq.~(\ref{moment-sum-smeared}) is absolutely convergent for arbitrary
$\mu$ when $z<1$.}
\label{moment-sum-smeared}
\end{equation}
Then, we obtain the light-cone amplitude ${\cal M}$ that corresponds to the 
distribution $\phi_V^\perp(x,\mu)$ by taking the limit of the sequence 
of ordinary functions that we use to define $\phi_V^\perp(x,\mu)$:
\begin{eqnarray}
\label{limit-eigenvalue-sum}
{\cal M}=\int_0^1 dx\, T_H(x,\mu)
\phi_{V}^\perp(x,\mu)&=&\lim_{z\to 1}
\int_0^1 dx\, T_H(x,\mu) \phi_S(x, z,\mu)
\nonumber \\
&=&
\lim_{z\to 1} \sum_{n=0}^\infty T_n(\mu) z^n \phi_{n}^\perp(\mu).
\end{eqnarray}
We note that  Eq.~(\ref{limit-eigenvalue-sum}) amounts to Abel
summation of the eigenfunction series. A mathematical proof of
Eq.~(\ref{limit-eigenvalue-sum}) is beyond the scope of this paper.
However, we will describe several numerical tests that strongly
support the validity of the Abel summation in
Eq.~(\ref{limit-eigenvalue-sum}).

In principle, one can use Eq.~(\ref{limit-eigenvalue-sum}) to compute
the light-cone amplitude, making use of Eq.~(\ref{moment-evolution}) to
take into account the scale evolution of the LCDA. In order to do this,
one would need carry out the sum in Eq.~(\ref{limit-eigenvalue-sum})
before taking limit $z\to 1$.  In practice, in carrying out a numerical
evaluation, one must include enough terms in the sum to guarantee that
the remainder is small for a given value of $|1-z|$. For the functions
$T_H(x,\mu)$ and $\phi_V^\perp(x,\mu)$ that we consider, this typically
requires that one include thousands of terms in order to achieve
percent-level precision.\footnote{We have verified numerically, for the
cases ${\cal M}^{(0,0)}$ and ${\cal M}^{(0,v^2)}$, with $\mu=m_Q$,
$m_H/2$, $m_H$, $2m_H$, $1$~TeV, and $2$~TeV, that the Abel
summation does converge, although very slowly, to the result that is
given by the Abel-Pad\'e procedure.}

A much more efficient procedure is to use Pad\'e approximants to
approximate the sum in Eq.~(\ref{limit-eigenvalue-sum}). As we have
mentioned, we refer to this method that makes use of a combination of
Abel summation and Pad\'e approximants as the Abel-Pad\'e method. The sum
in Eq.~(\ref{limit-eigenvalue-sum}) defines a function of $z$ that is
analytic for $|z|<1$. The Pad\'e approximant gives an approximate
analytic continuation of that function to larger values of $|z|$. In
particular, the Pad\'e approximant can give precise values of
Eq.~(\ref{limit-eigenvalue-sum}) for $z=1$, even when poles in the disc
$|z|<1$ render the radius of convergence of the series to be less than
$1$. Consequently, a Pad\'e-approximant expression that is based on a
given partial sum can give much better precision as $z\to 1$ than does
the original partial sum. For the functions $T_H(x,\mu)$ and
$\phi_V^\perp(x,\mu)$ that we consider, one can typically achieve much
better than percent-level precision by keeping $20$ terms in the partial
sum and generating a $10\times 10$ Pad\'e approximant.

In Appendix~\ref{sec:nonrel-num-tests}, we have tested the
Abel-Pad\'e method for the cases $\phi_V^\perp(x,\mu) \to
\phi_V^\perp(x,\mu_0)\to \delta^{(k)}(x-\tfrac{1}{2})$,
with $k=0,2,\ldots, 10$, and $T_H(x,\mu)\to T_H^{(0)}(x,\mu_0)$, {\it i.e.},
with no evolution. Analytic results are easily obtained in these cases,
and the Abel-Pad\'e expression converges quickly to them, even
though the eigenfunction series are not convergent for $k>0$. As can be
seen from the appendix of Ref.~\cite{Bodwin:2014bpa}, evolution of
$\phi_V^{\perp (0)}(x,\mu_0)$ to a higher scale generally improves that
convergence of the eigenfunction series. (This general property is
confirmed numerically in Appendix~\ref{sec:nonrel-num-tests}.) It seems,
therefore, that the zero-evolution tests of the Abel-Pad\'e method that we
have made are particularly demanding. We have also tested the
Abel-Pad\'e method by expanding the LL evolved expression for
$c_{2}(\mu)= f_V^\perp(\mu){\cal M}^{(0,v^2)}(\mu)$ as a series in
$\alpha_s$, using the Abel-Pad\'e method to compute the first
three terms in the series from their eigenfunction expansions (taking
$\mu_0=m_c\hbox{, }m_b$ and $\mu=m_H$), and comparing the results with
the analytic expressions for the first three terms in the series in
Eq.~(39b) of Ref.~\cite{Bodwin:2014bpa}.  Again, the Abel-Pad\'e
expressions converge rapidly to the analytic results, even though the
eigenfunction series themselves are not convergent.

We conclude that the Abel-Pad\'e method is reliable, and
we use it in this paper to sum all of the eigenvalue series for the
LCDAs.

\section{Comparison with a model LCDA \label{sec:model}}
 
In Ref.~\cite{Koenig:2015pha}, it was proposed to incorporate the
effects of the order-$v^2$ and order-$\alpha_s$ corrections to the
LCDA by making use of a model LCDA:
\begin{equation}
\phi_V^{\perp M}(x,\mu_0)=N_\sigma\frac{4x(1-x)}{\sqrt{2\pi}
\sigma_V(\mu_0)}
\exp\biggl[-\frac{(x-\tfrac{1}{2})^2}{2\sigma^{2}_V(\mu_0)}\biggr].
\label{KN-LCDA}
\end{equation}
Here, $N_\sigma$ is chosen so that 
\begin{equation}
\int_0^1 dx\, \phi_V^{\perp M}(x,\mu_0)=1.
\end{equation}
It is stated in Ref.~\cite{Koenig:2015pha} that the width parameter 
$\sigma_V(\mu_0)$ is chosen so that $\phi_V^{\perp{M}}(x,\mu_0)$ 
yields the second 
moment of $\phi_V^\perp(x,\mu)$ through linear order in $v^2$ and $\alpha_s$:
\begin{equation}
4\sigma_V^2(\mu_0)=\int_0^1 dx\, 
(2x-1)^2\phi_V^{\perp{M}}(x,\mu_0)\equiv \frac{\langle v^2\rangle_V}{3}+
\frac{C_F\alpha_s(\mu_0)}{4\pi}
\biggl(\frac{28}{9}-\frac{2}{3}\ln\frac{m_Q^2}{\mu_0^2}\biggr).
\label{KNLDME-2dmom}
\end{equation}
The initial scale is chosen to be $\mu_0=1$~GeV. 

The model LCDA circumvents the problem of the nonconvergence of the
eigenfunction series for ${\cal M}^{(0,v^2)}(\mu)$: Because
$\phi_V^{\perp{M}}(x,\mu_0)$ is an ordinary function of $x$, the
eigenfunction series converges. However, a number of assumptions go 
into the construction of the model LCDA. We now discuss the validity of 
those assumptions.

First, we note that the first equality in Eq.~(\ref{KNLDME-2dmom}) holds
only in the zero-width ($\sigma_V\to 0$) limit. In
Ref.~\cite{Koenig:2015pha}, numerical values of $\sigma_V(1~\hbox{GeV})$
were computed by equating $4\sigma_V^{2}$ to the expression on the right
side of the second equality in Eq.~(\ref{KNLDME-2dmom}). This procedure
leads to values for the second $x$ moments of
$\phi_V^{\perp{M}}(x,1~\hbox{GeV})$ that differ substantially from the
true values of second $x$ moments of $\phi_V^\perp(x,1~\hbox{GeV})$
through linear order in $v^2$ and $\alpha_s$. For example, in the case
of the $J/\psi$, with $m_c=1.4$~GeV and $\langle
v^2\rangle_{J/\psi}=0.225$, the second $x$ moment of
$\phi_V^{\perp{M}}(x,1~\hbox{GeV})$ is 0.120256, while the second
$x$ moment of $\phi_V^\perp(x,1~\hbox{GeV})$ through linear order in
$v^2$ and $\alpha_s$ is 0.207729. In fact, in this case, there is no
choice of $\sigma_V(1~\hbox{GeV})$ that yields the correct second
$x$ moment through linear order in $v^2$ and $\alpha_s$.

Second, we note that only the second $x$ moment of the order-$\alpha_s$
correction to the LCDA enters into the model LCDA. That is, there is an
implicit assumption that the order-$\alpha_s$ correction can be
adequately characterized by its second $x$ moment alone. However, the
order-$\alpha_s$ correction to the LCDA has substantial $x$ moments
beyond the second moment, and, so, this assumption seems to be
questionable. In contrast, only the second $x$ moment of the order-$v^2$
correction to the LCDA is nonvanishing.

Third, the functional form of the LCDA has implications for 
the higher $x$ moments of the LCDA. These higher $x$
moments are related to corrections to the LCDA of higher order in $v^2$
(see Refs.~\cite{Braguta:2007fja,Braguta:2007fh,Braguta:2012zza} and
Appendix~\ref{sec:nonrel-expansion}) and to higher $x$ moments of the
corrections to the LCDA of order $\alpha_s$ and higher. It is not clear
that the functional form of the LCDA accounts adequately for these
corrections. In Appendix~\ref{sec:nonrel-mom-sizes}, we examine $x$
moments of the model LCDA in order $\alpha_s^0$, using the relationships
between the $x$ moments of the LCDA and the NRQCD LDMEs that are given in
Refs.~\cite{Braguta:2007fja,Braguta:2007fh,Braguta:2012zza}. We find
that $x$ moments of the model LCDA are much larger than expectations
from the NRQCD velocity-scaling rules, suggesting that the model LCDA
leads to spuriously large corrections of higher order in
$v^2$.\footnote{\label{footnote:v-scaling}Strictly speaking, the
velocity-scaling rules state that an LDME $\langle v^n\rangle_V$, which
is defined by the obvious generalization of Eq.~(\ref{v2ME}), vanishes as
$v^n$ in the limit $v\to 0$. However, in phenomenology, the
velocity-scaling rules are usually taken to mean that $\langle
v^n\rangle_V$ is equal to $v^n$ times a coefficient of order 1.
This point of view is supported by the generalized Gremm-Kapustin
relation \cite{Bodwin:2006dn}.}

The ultimate test of the model LCDA is whether it leads to an accurate
numerical result for the light-cone amplitude. We will carry out such
a test by comparing the results for the light-cone amplitude that are
obtained from the model LCDA with the results for the light-cone
amplitude that are obtained from our calculation through orders
$\alpha_s$ and $v^2$. In doing so, we are implicitly assuming that the
expansions in the small parameters $\alpha_s$ and $v^2$ are valid and
that corrections beyond those in orders $\alpha_s$ and $v^2$ are small
in comparison with the corrections of orders $\alpha_s$ and $v^2$. One
could question whether the evolution from the scale $\mu_0$ to the scale
$m_H$ could invalidate the $\alpha_s$ and $v^2$ expansions. Regarding
the $\alpha_s$ expansion, evolution from the scale $1.0$~GeV to the
scale $m_H$ changes the order-$\alpha_s$ correction from $16\%$ of
the order-$\alpha_s^0$ contribution to $9\%$ of the
order-$\alpha_s^0$ contribution, suggesting that evolution does not
spoil the $\alpha_s$ expansion. Further tests of the $\alpha_s$
expansion would require the computation of corrections of still higher
orders in $\alpha_s$. We can investigate the convergence of the $v^2$
expansion (nonrelativistic expansion) and the effects of evolution on it
more completely, and we do so in Appendix~\ref{sec:nonrel-expansion}.
There, we test the numerical convergence of the nonrelativistic
expansion in order $\alpha_s^0$ for the example of the model LCDA.  We
find that the nonrelativistic expansion converges rapidly to the exact
result for the model LCDA at the scale $\mu=\mu_0$ and that it converges
even more rapidly at the scale $\mu=m_H$. The expansion through order
$v^2$ gives a good approximation to the exact result. We conclude that
the model LCDA, if it is valid, should not produce corrections beyond
the leading order in $\alpha_s$ and $v^2$ that deviate significantly
from the sum of the corrections of order $\alpha_s$ and order $v^2$ that
we compute in this paper.

We can assess whether the contributions of higher order that arise from
the model LCDA $\phi_V^{\perp{M}}(x,\mu)$ agree with the contributions
of order $v^2$ and order $\alpha_s$ that we compute by examining the
quantity
\begin{equation}
\Delta(\mu)=\frac{\alpha_s(\mu_0)}{4\pi}{\cal 
M}^{(0,1)}(\mu)+\langle v^2\rangle_V{\cal M}^{(0,v^2)}(\mu),
\end{equation}
where, in order to compare with $\phi_V^{\perp{M}}(x,\mu)$, we take 
$\mu_0=1$~GeV in  $\alpha_s(\mu_0)$ and, implicitly, in ${\cal 
M}^{(0,1)}(\mu)$ and ${\cal M}^{(0,v^2)}(\mu)$.
The equivalent expression for the model LCDA
$\phi_V^{\perp{M}}(x,\mu)$, is given, up to corrections of higher orders
in $\alpha_s$ and $v^2$, by
\begin{equation}
\Delta^{\!{M}}(\mu)=\int_0^1dx\, T_H^{(0)}(x,\mu) 
[\phi_V^{\perp{M}}(x,\mu)-\phi_V^{\perp (0)}(x,\mu)].
\end{equation}
In Table~\ref{tab:Delta-1GeV} we compare the values of
$\Delta(\mu_0)$ and $\Delta^{\!{M}}(\mu_0)$ for the 
$J/\psi$ and $\Upsilon(nS)$ states, using the values of the input 
parameters that are given in Ref.~\cite{Koenig:2015pha}. In the case of
$\Delta^{\!{M}}(\mu_0)$, we also show the values that result
from varying $\sigma_V(\mu_0)$ by $\pm 25\%$, as was suggested in
Ref.~\cite{Koenig:2015pha}.

\begin{table}[h]
\begin{center}
\begin{tabular}{lllll} 
\hline
$\phantom{xx}V\phantom{x}$ & 
$\phantom{x}\Delta(\mu_0)\phantom{^{M}\big|_{1.25\sigma_V}}$&
$\phantom{x}\Delta^{\!{M}}(\mu_0)_{\phantom{\sigma_V\to 1.25 \sigma_V}}$& 
$\phantom{x}\Delta^{\!{M}}(\mu_0)\big|_{\sigma_V\to 0.75\sigma_V}$&
$\phantom{x}\Delta^{\!{M}}(\mu_0)\big|_{\sigma_{V}\to 1.25\sigma_V}$
\\
\hline\hline
$\phantom{x}J/\psi\phantom{x}$ & 
\phantom{x}0.971375 & 
\phantom{x}0.843339 & 
\phantom{x}0.510365 &
\phantom{x}1.12087
\\
$\phantom{x}\Upsilon(1S)\phantom{x}$&
\phantom{x}0.0770658 &
\phantom{x}0.211269 &
\phantom{x}0.116175 &
\phantom{x}0.338490
\\
$\phantom{x}\Upsilon(2S)\phantom{x}$ & 
\phantom{x}0.209066 & 
\phantom{x}0.359150 & 
\phantom{x}0.195740 &
\phantom{x}0.563622
\\
$\phantom{x}\Upsilon(3S)\phantom{x}$ & 
\phantom{x}0.295732 &
\phantom{x}0.458135 &
\phantom{x}0.250834 &
\phantom{x}0.697510
\\
\hline
\end{tabular}
\caption{\label{tab:Delta-1GeV} Numerical values of $\Delta(\mu_0)$
and $\Delta^{\!{M}}(\mu_0)$ for $V=J/\psi$ and $\Upsilon(nS)$ at
$\mu_0=1~\textrm{GeV}$.  In the last two columns, we have
evaluated $\Delta^{\!{M}}(\mu_0)$ by replacing $\sigma_V(\mu_0)$ by $0.75$
and $1.25$ times its nominal value, respectively.}
\end{center}
\end{table}
\begin{table}[h]
\begin{center}
\begin{tabular}{lllll} 
\hline
$\phantom{xx}V\phantom{x}$ & 
$\phantom{x}\Delta(\mu)\phantom{^{M}\big|_{1.25\sigma_V}}$&
$\phantom{x}\Delta^{\!{M}}(\mu)_{\phantom{\sigma_V\to 1.25 \sigma_V}}$& 
$\phantom{x}\Delta^{\!{M}}(\mu)\big|_{\sigma_V\to 0.75\sigma_V}$&
$\phantom{x}\Delta^{\!{M}}(\mu)\big|_{\sigma_V\to 1.25\sigma_V}$
\\
\hline\hline
$\phantom{x}J/\psi$\phantom{x} & 
\phantom{x}0.684103 & 
\phantom{x}0.522962 & 
\phantom{x}0.337973 &
\phantom{x}0.666378
\\
$\phantom{x}\Upsilon(1S)\phantom{x}$&
\phantom{x}0.103008 &
\phantom{x}0.150110 &
\phantom{x}0.084148 &
\phantom{x}0.233466
\\
$\phantom{x}\Upsilon(2S)\phantom{x}$ & 
\phantom{x}0.200579 & 
\phantom{x}0.246479 & 
\phantom{x}0.139542 &
\phantom{x}0.368862
\\
$\phantom{x}\Upsilon(3S)\phantom{x}$ & 
\phantom{x}0.264641 &
\phantom{x}0.307054 &
\phantom{x}0.176647 &
\phantom{x}0.444124
\\
\hline
\end{tabular}
\caption{\label{tab:Delta-mH}
Numerical values of $\Delta(\mu)$ and $\Delta^{\!{M}}(\mu)$ for
$V=J/\psi$ and $\Upsilon(nS)$ at $\mu=m_H$. In the last two columns,
we have evaluated $\Delta^{\!{M}}(\mu)$ by replacing $\sigma_V(\mu_0)$
by $0.75$ and $1.25$ times its nominal value, respectively.
}
\end{center}
\end{table}
As can be seen from Table~\ref{tab:Delta-1GeV}, the central value of
$\Delta^{\!{M}}(\mu_0)$ deviates from the value of $\Delta(\mu_0)$
by $-13\%$ for the $J/\psi$, $+174\%$ for the $\Upsilon(1S)$, $+72\%$
for the $\Upsilon(2S)$, and $+55\%$ for the $\Upsilon(3S)$. We also see
that the result is very sensitive to the choice of
$\sigma_V(\mu_0)$: The values of $\Delta^{\!{M}}(\mu_0)$ vary
by factors of $2$ or more as $\sigma_V(\mu_0)$ is varied by $\pm
25\%$. [In contrast, $\Delta(\mu_0)$ would vary by less than $\pm 25\%$ if
the input parameter $\langle v^2\rangle_V$ were varied by $\pm 25\%$.]
Therefore, we regard the approximate agreement of the central value of
$\Delta^{\!{M}}(\mu_0)$ with the value of $\Delta(\mu_0)$
for the case of the $J/\psi$ as accidental.

In Table~\ref{tab:Delta-mH} we compare the values of $\Delta(m_H)$ and
$\Delta^{\!{M}}(m_H)$ for the $J/\psi$ and $\Upsilon(nS)$ states, using
the values of the input parameters at $1~\hbox{GeV}$ that are given in
Ref.~\cite{Koenig:2015pha}. Again, in the case of $\Delta^{\!{M}}(m_H)$, 
we also show the values that result from varying
$\sigma_V(\mu_0)$ by $\pm 25\%$. We make use of the
Abel-Pad\'e method in carrying out the evolution of $\mu$ from
$\mu_0=1~\hbox{GeV}$ to $m_H=125.09~\textrm{GeV}$, taking 100 terms in the
eigenfunction expansion and using a $50\times 50$ Pad\'e approximant.

In Ref.~\cite{Koenig:2015pha}, it was suggested that the evolution of
the model LCDA to the scale $\mu=m_H$ would reduce the dependence on the
specifics of the model. As can be seen from Table~\ref{tab:Delta-mH},
the central value of $\Delta^{\!{M}}(m_H)$ deviates from value of
$\Delta(m_H)$ by $-24\%$ for the $J/\psi$, $+46\%$ for the
$\Upsilon(1S)$, $+23\%$ for the $\Upsilon(2S)$, and $+16\%$ for the
$\Upsilon(3S)$. Comparison with Table~\ref{tab:Delta-1GeV} shows
that, in the case of the $J/\psi$, the deviation of
$\Delta^{\!{M}}(m_H)$ from $\Delta(m_H)$ actually increases as $\mu$ is
evolved from $\mu_0=1~\textrm{GeV}$ to $m_H$. 
While the deviations in the case of the
$\Upsilon(nS)$ states decrease as $\mu$ is evolved
from $1~\textrm{GeV}$ to
$m_H$, they are still rather large, especially in the case of the
$\Upsilon(1S)$. Furthermore, the results are very sensitive to
the choice of $\sigma_V(1~\hbox{GeV})$: The values of
$\Delta^{\!{M}}(m_H)$ vary by factors of $2$ or more as
$\sigma_V(1~\hbox{GeV})$ is varied by $\pm 25\%$.

We would expect the uncalculated corrections of higher orders in
$\alpha_s$ and $v^2$ to be of size $\alpha_s$ or $v^2$ relative to the
corrections that we have calculated. We see that the model LCDA of
Ref.~\cite{Koenig:2015pha} produces results that deviate from ours by
amounts that are much larger than the expected sizes of these
uncalculated corrections. Therefore, we conclude that the model LCDA of
Ref.~\cite{Koenig:2015pha} does not lead to reliable results for
contributions to the light-cone amplitude of the order-$\alpha_s$ and
order-$v^2$ corrections to the LCDA. However, because the value of
$\Delta(m_H)$ is small in comparison with the leading contribution to
the leading light-cone amplitude ${\cal M}^{(0,0)}=4$, the deviations of
$\Delta^{\!{M}}(m_H)$ from $\Delta(m_H)$ affect the light-cone amplitude
only at the level of about $4\%$ for the $J/\psi$ and at the level of
about $1\%$ for the $\Upsilon(nS)$ states.

\section{Computation of the decay rates \label{sec:computation}}
\subsection{Direct amplitude}

Our formula for the light-cone direct amplitude through order
$\alpha_s$, with NLL resummation of logarithms of $m_H^2/m_Q^2$, is
\begin{eqnarray}
i {\cal M}_{\rm dir}^{\rm LC}[\!\!&H&\to V+\gamma]
\nonumber \\
&=& \frac{i}{2} e e_Q \kappa_Q \overline{m}_Q(\mu) 
( \sqrt{2} G_F)^{1/2} 
\left( - \epsilon^*_V \cdot \epsilon^*_\gamma 
+ \frac{\epsilon^*_V \cdot p_\gamma p \cdot \epsilon_\gamma^*}
{p_\gamma \cdot p} \right) 
\frac{f_V^\perp (m_{H})}{f_V^\perp (\mu_0)} 
\frac{\sqrt{2 N_c} \sqrt{2 m_V}}{2 m_Q} \Psi_V(0)
\phantom{xxx}
\nonumber \\ && \times
\bigg\{
\left[ 1- \frac{5}{6} \langle v^2 \rangle_V + \frac{C_F\alpha_s (\mu_0)}{4 \pi}
\left( \log \frac{m_Q^2}{\mu_0^2} -8 \right) 
\right]
 {\cal M}^{(0,0)}(\mu)
\nonumber \\ && \hspace{5ex} 
+ 
\frac{\alpha_s (\mu)}{4 \pi} {\cal M}^{(1,0)}(\mu)+ 
\frac{\alpha_s (\mu_0)}{4 \pi}{\cal M}^{(0,1)}(\mu)+ 
\langle v^2 \rangle_V {\cal M}^{(0,v^2)}(\mu)
\bigg\},\label{A-dir}
\end{eqnarray}
where, in computing $i {\cal M}_{\rm dir}^{\rm LC} [H\to
V+\gamma]$, we take $e=\sqrt{4\pi\alpha(0)}$.

We note that the formula (\ref{A-dir}) does not contain any cross
terms of order $\alpha_s^2$, $\alpha_s v^2$, or $v^4$. In contrast, the
expressions in Ref.~\cite{Koenig:2015pha} do contain such cross terms
because the expansions of $T_H$ and the ratio $f_V^\perp/f_V$ in powers
of $\alpha_s$ and $\langle v^2\rangle_{V}$ appear as factors in the
expression that was used in Ref.~\cite{Koenig:2015pha} for the direct
amplitude. On the other hand, our computation contains cross terms that
arise from the ratio $f_V^\perp/f_V$ that are not contained in the
expression for $f_V^\perp/f_V$ in Ref.~\cite{Koenig:2015pha}. That is
because we use the values of the LDMEs that were extracted in
Refs.~\cite{Bodwin:2007fz,Chung:2010vz} by making use of a formula for
the quarkonium leptonic width that contains the expansion of the factor
$f_V$ in powers of $\alpha_s$ and $\langle v^2\rangle_{V}$. All of the
cross terms that we have mentioned appear at orders that are beyond the
claimed precision of our calculation or the calculation of
Ref.~\cite{Koenig:2015pha}. In our calculation, they are taken into
account in our estimates of uncertainties from uncalculated higher-order
corrections.

In the evolution of the expression in Eq.~(\ref{A-dir}), we choose
the initial scale to be $\mu_0 = m_Q$ and the final scale to be $\mu =
m_{H}$. This choice incorporates the logarithms of $m_H^2/m_Q^2$ into
the evolved expressions. We will discuss the effect of using the 
choice of scale $\mu_0=2m_Q$ in Sec.~\ref{sec:results}.

We note that, in Ref.~\cite{Koenig:2015pha}, the initial scales were
taken to be $1$~GeV for the LCDAs and $2$~GeV for the ratio of decay
constants $f_V^\perp/f_V$. This latter choice is somewhat inconsistent
with the use of values of $\langle v^2\rangle_{V}$ from
Refs.~\cite{Bodwin:2007fz,Chung:2010vz}, as they were extracted by
making use of the expansion of $f_V$ in powers of $\alpha_s$ and
$\langle v^2\rangle_{V}$, with $\alpha_s(\mu)$ evaluated at the scale
$m_V$.

\subsection{Indirect amplitude}

In computing the indirect amplitude, we follow 
Refs.~\cite{Bodwin:2013gca,Bodwin:2014bpa}, taking
\begin{subequations}
\begin{equation}
i{\cal M}_{\rm ind}=i{\cal A}_{\rm ind}
\left(-\epsilon_V^*\cdot \epsilon^*_{\gamma}
+\frac{\epsilon_V^*\cdot p_\gamma  \,
p_V\cdot\epsilon^*_{\gamma}}
{p_\gamma\cdot p_V}\right),
\end{equation}
where 
\begin{equation}
{\cal A}_{\rm ind}=\frac{g_{V\gamma}\sqrt{4\pi\alpha(m_V)m_H}}{m_V^2}
\bigg[16\pi\frac{\alpha(m_V)}{\alpha(0)}
\Gamma(H\to\gamma\gamma)\bigg]^{\frac{1}{2}},
\label{Aind}
\end{equation}
and $g_{V\gamma}$ is expressed in terms of the width of $V$ into leptons 
\cite{Bodwin:2013gca}:
\begin{equation}
g_{V\gamma}=-
\frac{e_Q}{|e_Q|}\bigg[\frac{3m_V^3\Gamma(V\to 
\ell^+\ell^-)}{4\pi \alpha^2(m_V)}\bigg]^{\frac{1}{2}}.
\label{gVgam}
\end{equation}
\end{subequations}
We obtain $\Gamma(H\to \gamma\gamma)$ from the values of the Higgs-boson
total width and branching fraction to $\gamma\gamma$ in
Refs.~\cite{Dittmaier:2011ti,Dittmaier:2012vm}. In the expression
(\ref{Aind}) for ${\cal A}_{\rm ind}$, we neglect a small phase that is
about 0.005. As in Ref.~\cite{Bodwin:2013gca}, we have chosen the scales
of the electromagnetic coupling as follows: we use $\alpha(m_V)$ to
compute $g_{V\gamma}$ from the $V$ leptonic width, we use $e=\sqrt{4\pi
\alpha(m_V)}$ for the couplings of the virtual photon, and we use
$e=\sqrt{4\pi \alpha(0)}$ for the coupling of the real photon. We have
also compensated for the fact that $\Gamma(H\to\gamma\gamma)$ was
computed in Refs.~\cite{Dittmaier:2011ti,Dittmaier:2012vm} using
$e=\sqrt{4\pi \alpha(0)}$.

In contrast with the calculations in
Refs.~\cite{Bodwin:2013gca,Koenig:2015pha}, our calculation of
$\cal{A}_{\rm ind}$ does not include contributions that are
suppressed as $m_V^2$ divided by combinations of $m_H^2$, $m_t^2$,
$m_Z^2$, or $m_W^2$, where $m_t$, $m_W$, and $m_Z$ are the masses of
the top quark, $W^{\pm}$ boson, and $Z^0$ boson, respectively.
Such contributions can arise not only from explicit $m_V$ terms in the
amplitude for $H\to \gamma\gamma^*$, but also from electroweak
corrections to the amplitude for $H\to V+\gamma$. In the latter, it is
not possible to distinguish between direct and indirect processes in a
gauge-invariant way.

\subsection{Numerical inputs}

We take the pole masses to be the one-loop values $m_c=1.483$~GeV and
$m_b=4.580$~GeV, we take the $\overline{\rm MS}$ masses to be
$\overline{m}_c=1.275$~GeV and $\overline{m}_b=4.18$~GeV, and we take
$m_H=125.09\pm 0.21\hbox{ (stat.)}\pm 0.11\hbox{ (syst.)}$~GeV, which
implies, from the tables in
Refs.~~\cite{Dittmaier:2011ti,Dittmaier:2012vm}, that $\Gamma(H\to
\gamma \gamma)=(9.308\pm 0.120)\times 10^{-6}$~GeV. Here, we
have included a 1\% uncertainty from uncalculated higher-order terms in
the theoretical expression, an uncertainty of $0.022\%$ from the
uncertainty in $m_t$, an uncertainty of $0.024\%$ from the uncertainty
in $m_W$, and an uncertainty of $0.82\%$ from the uncertainty in $m_H$.
Our values for $|\Psi_V(0)|^2$ and $\langle v^2\rangle_V$ are shown in
Table~\ref{num-inputs}. Following Ref.~\cite{Bodwin:2013gca}, we use the
values from Refs.~\cite{Bodwin:2007fz,Chung:2010vz}, except that
\begin{table}[h]
\begin{center}
\begin{tabular}{lll} 
\hline
$\phantom{xx}V\phantom{x}$ & 
$\phantom{x}|\Psi_V(0)|^2~(\textrm{GeV}^3)$&
$\phantom{xxxxxxx}\langle v^2\rangle_V$
\\
\hline\hline
$\phantom{x}J/\psi$\phantom{x} & 
$\phantom{-}0.0729\pm 0.0109$ & 
$\phantom{xx}\phantom{-}0.201\pm 0.064$ 
\\
$\phantom{x}\Upsilon(1S)\phantom{xx}$&
$\phantom{-}0.512\pm 0.035$ &
$\phantom{x}-0.00920\pm 0.0105$
\\
$\phantom{x}\Upsilon(2S)\phantom{x}$ & 
$\phantom{-}0.271\pm 0.019$ & 
$\phantom{xx}\phantom{-} 0.0905\pm 0.0109$
\\
$\phantom{x}\Upsilon(3S)\phantom{x}$ & 
$\phantom{-}0.213\pm 0.015$ &
$\phantom{xx}\phantom{-}0.157\pm 0.017$
\\
\hline
\end{tabular}
\caption{\label{num-inputs}
Values of $|\Psi_V(0)|^2$ in units of $\textrm{GeV}^3$ and $\langle
v^2\rangle_V$ for $V=J/\psi$ and $\Upsilon(nS)$. These values have been
taken from Refs.~\cite{Bodwin:2007fz,Chung:2010vz}, except for the
uncertainties in $\langle v^2\rangle_{\Upsilon(1S)}$ and $\langle
v^2\rangle_{\Upsilon(2S)}$, which are described in the text.}
\end{center}
\end{table}
we have increased the uncertainties in
$\langle v^2\rangle_{\Upsilon(1S)}$ and $\langle
v^2\rangle_{\Upsilon(2S)}$ from those in Ref.~\cite{Chung:2010vz}. The
uncertainty from uncalculated corrections of order $v^4$ was estimated
in Ref.~\cite{Chung:2010vz} by multiplying the central value of $\langle
v^2\rangle_{\Upsilon(nS)}$ by $v^2$, where $v^2=0.1$ was used for the
$\Upsilon(nS)$ states. Because the central value of $\langle
v^2\rangle_{\Upsilon(1S)}$ is anomalously small (much less than $v^2$),
owing to an accidental cancellation in the $\overline{\rm MS}$
subtraction scheme, the estimate of the uncalculated order-$v^4$
corrections in Ref.~\cite{Chung:2010vz} considerably understates the
uncertainty from this source. The uncertainty for $\langle
v^2\rangle_{\Upsilon(2S)}$ was also slightly underestimated. Instead of
using the estimates in Ref.~\cite{Chung:2010vz}, we take the
uncertainties in $\langle v^2\rangle_{\Upsilon(1S)}$ and $\langle
v^2\rangle_{\Upsilon(2S)}$ from uncalculated order-$v^4$ corrections to
be $v^4=0.01$.

\subsection{Sources of uncertainties}

In calculating the decay rates, we take into account uncertainties in
both the direct and indirect amplitudes, as is described below. In
computing branching fractions, we also take into account the uncertainty
in the total decay width of the Higgs boson
\cite{Dittmaier:2011ti,Dittmaier:2012vm}.

\subsubsection{Direct amplitude}

In the direct amplitude, we include the uncertainties that arise from
the uncertainties in $\Psi_V(0)$ and the uncertainties in $\langle
v^2\rangle_{V}$.  We also include the uncertainties that arise from
uncalculated corrections of order $\alpha_s^2$, order $\alpha_s v^2$, and
order $v^4$. We estimate the uncertainties from these uncalculated
corrections, relative to the lowest nontrivial order in the direct
amplitude, to be $\{[C_F C_A\alpha_s^2(m_Q)/\pi^2]^2+[C_F\alpha_s(m_Q)
v^2/\pi]^2+[v^4]^2\}^{1/2}$ for the real part of the direct amplitude and
$\{[C_A\alpha_s(m_Q)/\pi]^2+[v^2]^2\}^{1/2}$ for the imaginary part of the
direct amplitude. (Note that the real part of the direct amplitude
starts in absolute order $\alpha_s^0$ and the imaginary part of the
direct amplitude starts in absolute order $\alpha_s$.) We take $v^2=0.3$
for the $J/\psi$ and $v^2=0.1$ for the $\Upsilon(nS)$ states.

In Ref.~\cite{Koenig:2015pha}, it was suggested that the uncertainties
in $\Psi_V(0)$ and $\langle v^2\rangle_{V}$ were underestimated in
Refs.~\cite{Bodwin:2007fz,Chung:2010vz}. We now address these 
issues.

One difficulty that was raised in Ref.~\cite{Koenig:2015pha} is that
one-loop pole masses were used in
Refs.~\cite{Bodwin:2007fz,Chung:2010vz} in the one-loop expression for
$\Gamma(V\to \ell^+\ell^-)$, which was used to compute $\Psi_V(0)$. The
objection is that the pole mass is ill defined outside of perturbation
theory and is subject to renormalon ambiguities. However, in
Refs.~\cite{Bodwin:2007fz,Chung:2010vz}, the pole mass was used in
conjunction with one-loop corrections to $\Gamma(V\to \ell^+\ell^-)$
that are calculated using the pole mass. This is equivalent, up to
corrections of higher order in $\alpha_s$, to the use of the
$\overline{\rm MS}$ mass in conjunction with one-loop corrections to
$\Gamma(V\to \ell^+\ell^-)$ that are calculated using the $\overline{\rm
MS}$ mass. At one-loop order, the numerical difference between the two
procedures is small.

Another difficulty that was raised in Ref.~\cite{Koenig:2015pha} is that
the perturbation series for $\Gamma(V\to \ell^+\ell^-)$ has very large
corrections at two-loop and three-loop orders
\cite{Beneke:1997jm,Czarnecki:2001zc,Marquard:2014pea,Beneke:2014qea}.
The perturbation series was truncated at one-loop order in
Refs.~\cite{Bodwin:2007fz,Chung:2010vz}. While an understanding of the
large two-loop and three-loop corrections to  $\Gamma(V\to
\ell^+\ell^-)$ is still lacking, it should be noted that the analyses in
Refs.~\cite{Bodwin:2007fz,Chung:2010vz} of the wave functions at the
origin for the vector states $V$ and the pseudoscalar states $P$,
which make use of the one-loop expressions for $\Gamma(V\to
\ell^+\ell^-)$ and $\Gamma(P\to \gamma\gamma)$, result in the same
values for the corresponding $V$ and $P$ wave functions at the
origin, up to differences whose numerical sizes are of order $v^2$,
in agreement with NRQCD velocity scaling. This agreement was
obtained in spite of the fact that both $\Gamma(V\to \ell^+\ell^-)$ and
$\Gamma(P\to \gamma\gamma)$ receive different large corrections in
two-loop order \cite{Czarnecki:2001zc}, and it suggests that one-loop
truncation is a reasonable procedure at the current level of precision.

In Ref.~\cite{Koenig:2015pha}, the ratio $f_V^\perp(\mu)/f_V$ appears,
where the direct amplitude is proportional to $f_V^\perp(\mu)$ and
$\Gamma(V\to \ell^+\ell^-)$ is proportional to $f_V^2$. The expression
for this ratio through order $\alpha_s$ (one-loop order) and through
order $v^2$ was used in Ref.~\cite{Koenig:2015pha}, rather than
the separate expressions for the numerator and the denominator. At the
one-loop order, for which the perturbation series for the numerator and
the denominator are separately well behaved, the use of the ratio
confers no particular advantage. At the two-loop order, at which the
perturbation series for $\Gamma(V\to \ell^+\ell^-)\propto f_V^2$ is
badly behaved, the ratio could conceivably be better behaved than either
the numerator or the denominator. However, this conjecture has not yet
been validated, as the two-loop corrections to $f_V^\perp(\mu)$ have
yet to be calculated.

Finally, we mention that, even if we assume that the uncertainty in
the perturbative expression for $\Gamma(V\to \ell^+\ell^-)$ is as large as
$100\%$ of the contribution of the one-loop term, the resulting
uncertainty in $\langle v^2\rangle_{V}$ is comparable to that from other
sources of uncertainty. If we repeat the analyses of
Refs.~\cite{Bodwin:2007fz,Chung:2010vz}, but allow the  perturbative
expression for $\Gamma(V\to \ell^+\ell^-)$ to vary by $100\%$ of the
contribution of the one-loop term, then the values for $\langle         
v^2\rangle_{V}$ deviate from the central value by a maximum of $88\%$,
$143\%$, $62\%$, and $135\%$ of the error bars in Table~\ref{num-inputs}
for the $J/\psi$, $\Upsilon(1S)$, $\Upsilon(2S)$, and $\Upsilon(3S)$,
respectively. 
Hence, the uncertainties in $\langle v^2\rangle_{V}$
that are given in Table~\ref{num-inputs} seem to be ample to take into
account the uncertainties in the perturbative expression for $\Gamma(V\to
\ell^+\ell^-)$.

\subsubsection{Indirect amplitude}

In estimating the uncertainties in the indirect amplitude, we follow the
method that is given in footnote~2 of Ref.~\cite{Bodwin:2013gca}. As
we have already mentioned, we include in $\Gamma(H\to \gamma \gamma)$
the uncertainties that arise from uncalculated higher-order terms in the
theoretical expression, the uncertainty in $m_t$, the uncertainty in
$m_W$, and the uncertainty in $m_H$. We assume that the uncertainties
in the leptonic decay widths are 2.5\% for the $J/\psi$, 1.3\% for the
$\Upsilon(1S)$, and 1.8\% for the $\Upsilon(2S)$ and $\Upsilon(3S)$
states. We take the relative uncertainty in the indirect amplitude
from uncalculated mass corrections to be $m_V^2/m_H^2$.

\subsection{Method for computing uncertainties in the decay rates 
\label{sec:computing-uncertainties}}

Owing to cancellations between the direct and indirect amplitudes, small
variations in those amplitudes can result in very nonlinear changes
in $\Gamma(H\to V+\gamma)$. Hence, one cannot reliably estimate the
total uncertainty in $\Gamma(H\to V+\gamma)$ simply by adding the
uncertainties from the individual sources in quadrature. Instead, we use
the following method to estimate the total uncertainty in $\Gamma(H\to
V+\gamma)$. We write $\Gamma(H\to V+\gamma)$ as a function of the
various uncertain input parameters and the normalizations of the
direct and indirect amplitudes. Then, we find the global maximum and
global minimum of $\Gamma(H\to V+\gamma)$ in a region about the central
values of the input parameters and normalizations that is
constrained as
\begin{equation}
\sum_i \left|\frac{c_i-c_{i0}}{\Delta c_i}\right|^2\le 1,
\end{equation}
where the $c_i$ are the input parameters and normalizations, 
the $c_{i0}$ are the central values of the $c_i$, and the $\Delta c_i$ are
the uncertainties in the $c_i$. We take the upper (lower) error bar on
$\Gamma(H\to V+\gamma)$ to be the global maximum (minimum) of
$\Gamma(H\to V+\gamma)$ minus the central value of $\Gamma(H\to
V+\gamma)$.

\section{Results\label{sec:results}}

Our results for the direct and indirect amplitudes are given in
Table~\ref{num-amplitudes}, where the evolution of the direct amplitudes
has been computed by the Abel-Pad\'e method, and we have
retained $100$ terms in the eigenvalues series and used $50\times 50$
Pad\'e approximants.

We note that, had we made the choice of initial scale $\mu_0=2m_Q$, that
would have shifted our results for the real parts of the direct
amplitudes by $+13\%$, $+4\%$, $+4\%$, and $+4\%$ for the $J/\psi$, 
$\Upsilon(1S)$, $\Upsilon(2S)$, and $\Upsilon(3S)$, respectively.
These shifts are within our estimated uncertainties for the real parts
of the direct amplitudes, which are $15\%$, $4\%$, $4\%$, and $4\%$ for
the $J/\psi$, $\Upsilon(1S)$, $\Upsilon(2S)$, and $\Upsilon(3S)$,
respectively. The choice of initial scale $\mu_0=2m_Q$ would have
shifted our results for the imaginary parts of the direct amplitudes by
$+0.1\%$ and $-1.6\%$ for the $J/\psi$ and $\Upsilon(nS)$ states,
respectively. These shifts are well within our estimated
uncertainties for the imaginary parts of the direct amplitudes.

The results in Ref.~\cite{Bodwin:2014bpa} for the real parts of the
direct amplitudes are considerably larger than our results, by $66\%$,
$20\%$, $22\%$, and $23\%$ for the $J/\psi$, $\Upsilon(1S)$,
$\Upsilon(2S)$, and $\Upsilon(3S)$, respectively. These differences are
due, primarily, to the use of LL evolution, rather than NLL evolution,
for $\overline{m}(\mu)$ and $f_V^\perp(\mu)$ in
Ref.~\cite{Bodwin:2014bpa}. The differences are larger than the values
that one obtains simply by considering the generic size of a
next-to-leading logarithm, namely,
$[\alpha_s(m_Q)/\pi]^2\log(m_H^2/m_Q^2)$. In the case of
$\phi_V^\perp(x,\mu)$, the use of NLL evolution, rather than LL
evolution, changes the direct amplitude by about 0.12\% for the $J/\psi$
and about 0.16--0.17\% for the $\Upsilon(nS)$ states. These changes are
negligible in comparison with the uncertainties in the direct
amplitudes. The use of the Abel-Pad\'e method to sum the
logarithms of $c_2(\mu)=f_V^\perp(\mu){\cal M}^{(0,v^2)}(\mu)$ to all
orders in $\alpha_s$, rather than through order $\alpha_s^2$, as in
Ref.~\cite{Bodwin:2014bpa}, amounts to about a $10\%$ change in the case
of the $J/\psi$ and to about a $4\%$ change in the case of the
$\Upsilon(nS)$ states. Since the corrections to the direct amplitude
that arise from $c_2(\mu)$ are about $4\%$ in the case of the $J/\psi$
and about $3\%$ in the case of the $\Upsilon(nS)$ states, the changes to
the direct amplitude that result from the use of the
Abel-Pad\'e method are negligible in comparison to the
uncertainties.

The results in Ref.~\cite{Koenig:2015pha} for the ratio of the real part
of the direct amplitude to the indirect amplitude are slightly larger
than our results for that ratio, by $17\%$, $7\%$, $7\%$, and $8.5\%$
for the $J/\psi$, $\Upsilon(1S)$, $\Upsilon(2S)$, and $\Upsilon(3S)$,
respectively. These differences are somewhat larger than our relative
uncertainties in the real parts of the direct amplitudes, and they are also
larger than the uncertainties that are given in
Ref.~\cite{Koenig:2015pha} for the ratio of the real part of the direct
amplitude to the indirect amplitude. 

The results in Ref.~\cite{Koenig:2015pha} for the ratio of the imaginary
part of the direct amplitude to the indirect amplitude differ from our
results for that ratio by $-12\%$, $9\%$, $4\%$, and $1\%$ for the
$J/\psi$, $\Upsilon(1S)$, $\Upsilon(2S)$, and $\Upsilon(3S)$,
respectively. These differences are well within our relative
uncertainties for the imaginary parts of the direct amplitudes.

As we have already mentioned, there are several possible 
sources of these differences between our results for the direct 
amplitudes and those of Ref.~\cite{Koenig:2015pha}. (1) 
Our initial scales for the evolution of $f_V^\perp(\mu)$ and the LCDAs 
are different from those in Ref.~\cite{Koenig:2015pha}. (2) Our formula for 
the direct amplitude (\ref{A-dir}) treats cross terms of order 
$\alpha_s^2$, $\alpha_s v^2$, and $v^4$ differently than does the 
corresponding formula in Ref.~\cite{Koenig:2015pha}. (3) Our treatment 
of the order $\alpha_s$ and order $v^2$ corrections to the LCDA is 
different from the model-LCDA treatment of Ref.~\cite{Koenig:2015pha}.

\begin{table}[h]
\begin{center}
\begin{tabular}{lll} 
\hline
$\phantom{xx}V\phantom{x}$ & 
$\phantom{x}\alpha_V$&
$\phantom{x}\beta_V$
\\
\hline\hline
$\phantom{x}J/\psi$& 
$11.71 \pm 0.16$ & 
$(0.627_{-0.094}^{+0.092}) + 
(0.118_{-0.054}^{+0.054} ) i$ 
\\
$\phantom{x}\Upsilon(1S)\phantom{xx}$&
$3.283 \pm 0.035\phantom{xxxx}$ &
$(2.908^{+0.122}_{-0.124}) + (0.391^{+0.092}_{-0.092}) i$
\\
$\phantom{x}\Upsilon(2S)$ & 
$2.155 \pm 0.028$ & 
$(2.036^{+0.087}_{-0.089} )
+ (0.293^{+0.069}_{-0.069} ) i$
\\
$\phantom{x}\Upsilon(3S)$ & 
$1.803 \pm 0.023$ &
$(1.749 ^{+0.077}_{-0.078})
+ (0.264^{+0.062}_{-0.062}) i$
\\
\hline
\end{tabular}
\caption{\label{num-amplitudes}
Values of the parameters $\alpha_V$ and $\beta_V$ in $\Gamma(H\to
V+\gamma)=|\alpha_V-\beta_V\kappa_Q|^{2}\times 10^{-10}~\textrm{GeV}$
for $V=J/\psi$ and $\Upsilon(nS)$.
}
\end{center}
\end{table}

Our results for the SM decay rates and branching fractions
($\kappa_Q=1$) are given in Table~\ref{tab:num-rates}. In computing
the uncertainties in the branching fractions, we have included the
effect of the uncertainty in the Higgs-boson total width. 

\begin{table}[h]
\begin{center}
\begin{tabular}{lll} 
\hline
$\phantom{xx}V\phantom{x}$ & 
$\phantom{x}\Gamma(H\to V+\gamma)~(\textrm{GeV})$&
$\phantom{xx}{\rm Br}(H\to V+\gamma)$
\\
\hline\hline
$\phantom{x}J/\psi$\phantom{x} & 
$\phantom{x}1.228^{+0.042}_{-0.042} \times 10^{-8} $ & 
$\phantom{xx}3.01^{+0.16}_{-0.15} \times 10^{-6}$ 
\\
$\phantom{x}\Upsilon(1S)\phantom{xx}$&
$\phantom{x}2.94^{+1.25}_{-1.02} \times 10^{-11}\phantom{xx}$ &
$\phantom{xx}7.19^{+3.07}_{-2.52} \times 10^{-9}$
\\
$\phantom{x}\Upsilon(2S)\phantom{x}$ & 
$\phantom{x}1.00^{+0.48}_{-0.39} \times 10^{-11}$ & 
$\phantom{xx}2.45^{+1.18}_{-0.96} \times 10^{-9}$
\\
$\phantom{x}\Upsilon(3S)\phantom{x}$ & 
$\phantom{x}7.27^{+3.67}_{-2.93} \times 10^{-12}$ &
$\phantom{xx}1.78^{+0.90}_{-0.72} \times 10^{-9}$
\\
\hline
\end{tabular}
\caption{\label{tab:num-rates}
SM values of $\Gamma(H\to V+\gamma)$ in units of GeV and ${\rm Br}(H\to V+\gamma)$
for $V=J/\psi$ and $\Upsilon(nS)$.
}
\end{center}
\end{table}

Our results for the SM decay rates agree with those in
Ref.~\cite{Bodwin:2014bpa}, within the uncertainties that are given in
Ref.~\cite{Bodwin:2014bpa}, except in the case of the $\Upsilon(1S)$. In
this case, the real parts of the SM direct and indirect amplitudes
nearly cancel, and so, as was pointed out in
Ref.~\cite{Koenig:2015pha}, the inclusion of the imaginary part of the
direct amplitude results in a significant increase in the rate.

Our results for the SM branching fractions agree with those in
Ref.~\cite{Koenig:2015pha}, within our uncertainties. Note that our
estimated uncertainties in the branching fractions are comparable to
those of Ref.~\cite{Koenig:2015pha}, except in the case of the
$\Upsilon(1S)$, for which our uncertainty is considerably larger. Since,
in the $\Upsilon(1S)$ case, our uncertainty in the ratio of the direct
amplitude to the indirect amplitude is essentially the same as
Ref.~\cite{Koenig:2015pha}, we suspect that the difference between the
uncertainty estimates arises because of the highly nonlinear dependences
of the decay rate on the input parameters. (See
Sec.~\ref{sec:computing-uncertainties}.)

\section{Summary and discussion \label{sec:summary}}

In this paper, we have presented new calculations of Higgs-boson decay
rates to vector heavy-quarkonium states plus a photon, where we have
considered the vector quarkonium states $J/\psi$ and $\Upsilon(nS)$,
with $n=$ 1, 2, or 3. As was pointed out in Ref.~\cite{Bodwin:2013gca},
these decay rates, when compared with data from a high-luminosity LHC
run, can provide information about the $Hc\bar c$ and $Hb\bar b$
couplings. Our calculation is carried out in the light-cone formalism in
which the nonperturbative parts of the quarkonium 
LCDAs are expressed in terms of NRQCD
long-distance matrix elements \cite{Wang:2013ywc}. Our calculations of
the direct decay amplitudes take into account corrections through
order $\alpha_s$ and order $v^2$ and include resummations of
logarithms of $m_H^2/m_Q^2$ to all orders in $\alpha_s$ through order 
$v^2$ at NLL accuracy.

In order to resum logarithms that are associated with the quarkonium
LCDAs, we have devised a new method, called the Abel-Pad\'e method, which
makes use of Abel summation, accelerated through the use of Pad\'e
approximants. The new method allows us to compute formally divergent
sums over the eigenfunctions of the LO evolution kernels. These
divergences arise because the LCDAs at initial scale of the evolution
are generalized functions (distributions) of the light-cone fractions,
rather than ordinary functions. The Abel-Pad\'e method defines these
distributions as sequences of ordinary functions and, hence, gives
finite and unambiguous results for the formally divergent sums. We have
tested this method numerically against known analytic results for the
LCDAs, and we find that it converges quickly and reliably to the values
from analytic calculations. It solves the general problem of carrying
out the scale evolution in a nonrelativistic expansion of the LCDA for
heavy-quarkonium systems, and it should be applicable in other
situations in which series of orthogonal polynomials fail to converge
when they are used to represent generalized functions. Using the 
Abel-Pad\'e method, we were able to make definitive calculations of the
LCDA-evolution effects in Higgs-boson decays to a quarkonium plus a
photon.

We have compared the Abel-Pad\'e method with the approach of
Ref.~\cite{Koenig:2015pha}, in which a model LCDA is used to take into
account relativistic and QCD corrections to the LCDA. In contrast with
the model approach, the Abel-Pad\'e method makes use only of the
calculated nonrelativistic corrections \cite{Bodwin:2014bpa} and
QCD corrections \cite{Wang:2013ywc}, and does not introduce any new
model assumptions. We find that the model of Ref.~\cite{Koenig:2015pha}
gives results that disagree substantially with those from the
Abel-Pad\'e method and that the model results are very sensitive
to the choices of model parameters. It turns out that the
relativistic and QCD corrections to the LCDA have only small effects
on the direct decay amplitude, and so the large differences between the
model and Abel-Pad\'e calculations of the relativistic and QCD
corrections to the LCDA have only small effects on the
decay rates.

Our results for the ratios of the direct decay amplitudes to the
indirect decay amplitudes are in reasonable agreement with those in
Ref.~\cite{Koenig:2015pha}. Since the indirect decay amplitude can be
determined quite precisely, this implies that our direct decay
amplitudes are in reasonable  agreement with those in
Ref.~\cite{Koenig:2015pha}. Our results for the real parts of the direct
decay amplitudes are considerably smaller than those in
Ref.~\cite{Bodwin:2014bpa}, owing to the use in
Ref.~\cite{Bodwin:2014bpa} of LL resummation, rather than NLL
resummation, of the logarithms of $m_H^2/m_Q^2$. Our result implies
that the sensitivities of the decay rates to the $HQ\bar Q$ couplings
are considerably smaller than the sensitivities that were suggested in
Ref.~\cite{Bodwin:2014bpa}, especially in the case of the $J/\psi$.

Our results for the SM decay rates are in good agreement with those of
Ref.~\cite{Bodwin:2014bpa}, except in the case of the $\Upsilon(1S)$. As
was pointed out in Ref.~\cite{Koenig:2015pha}, it is important to
include the imaginary part of the direct amplitude in the case of the
decay to $\Upsilon(1S)$ because there is an almost exact cancellation
between the real parts of the direct and indirect amplitudes. The
inclusion of the imaginary part of the direct amplitude in our
calculation increases the decay rate in the $\Upsilon(1S)$ case
substantially in comparison to the rate that is given in
Ref.~\cite{Bodwin:2014bpa}.

The branching fractions that we find are in good agreement with those in
Ref.~\cite{Koenig:2015pha}. Our uncertainty estimate in the case of the
$\Upsilon(1S)$ differs from that in Ref.~\cite{Koenig:2015pha}, possibly
owing to the highly nonlinear dependence of the rate on the input
parameters. In Sec.~\ref{sec:computing-uncertainties}, we have presented
a novel method for estimating the uncertainties in the presence of such
nonlinearities.

In the calculations that we have described, there is one important
theoretical issue that remains unresolved. The direct amplitude is
proportional to the quarkonium wave function at the origin. The wave
function at the origin is usually determined by comparing the
theoretical expression for the quarkonium decay rate to leptons with the
measured rate. In Refs.~\cite{Bodwin:2014bpa,Koenig:2015pha}, and in the
present work, the one-loop expression for the decay rate was used. Two-
and three-loop expressions exist
\cite{Beneke:1997jm,Czarnecki:2001zc,Beneke:2014qea}, but the
higher-loop corrections apparently destroy the convergence of the
perturbation series. As we have mentioned, the one-loop analyses in
Refs.~\cite{Bodwin:2007fz,Chung:2010vz} result in values for the
corresponding vector and pseudoscalar wave functions at the origin that
agree, up to differences whose numerical sizes are of relative order
$v^2$. This agreement, which is predicted by the NRQCD
velocity-scaling rules, is obtained in spite of the fact that the
two-loop corrections to the vector decays to leptons and the pseudoscalar
decays to two photons are large and different in relative size. The
agreement suggests that the one-loop truncations of the perturbation
series may lead to reasonable results for the wave functions at the
origin at a level of precision of order $v^2$.

In Ref.~\cite{Koenig:2015pha}, the ratio of decay constants
$f_V^\perp/f_V$ appears. The direct $H\to V+\gamma$ amplitude is
proportional to $f_V^\perp$, and the leptonic width of the vector
quarkonium is proportional to $f_V^2$. This ratio is evaluated through
order $\alpha_s$ (one-loop order) and order $v^2$. Hence, the
calculation in Ref.~\cite{Koenig:2015pha} also truncates the
perturbation series for the leptonic width at one-loop level. It is
conceivable that the ratio $f_V^\perp/f_V$ is better behaved than
either the numerator or the denominator. A calculation of two-loop QCD
corrections to $f_V^\perp$ would help to test this conjecture.

Higgs-boson decays to a vector quarkonium plus a photon provide
important opportunities to measure the $HQ\bar Q$ couplings at the LHC
and are the only known processes that can provide phase information
about those couplings. In order to take advantage of these opportunities
to determine the $HQ\bar Q$ couplings, it is essential to have the
theoretical calculations of the decay rates under good control. In this
paper, we have addressed the issue of the divergences that appear when
one uses conventional eigenfunction-expansion methods to resum the
logarithms of $m_H^2/m_Q^2$ that appear in the nonrelativistic expansions
of the quarkonium light-cone distribution amplitudes. With the
resolution of this issue, we believe that, aside from the matter of the 
determination of quarkonium wave functions at the origin that we have
mentioned above, calculations of the rates for Higgs-boson decays to
vector quarkonia plus a photon are now on a sound theoretical footing.

\appendix
\section{Evolution of the running mass and decay 
constant\label{sec:mass-decayconst-evolve}}

Here, we collect formulas at NLL accuracy for the evolution of the 
running $\overline{\rm MS}$ mass $\overline{m}(\mu)$ 
\cite{Tarrach:1980up}
and the decay constant $f_V^\perp (\mu)$
\cite{Broadhurst:1994se}:
\begin{subequations}
\begin{eqnarray}
\frac{\overline{m}(\mu)}{\overline{m}(\mu_0)} 
&=& 
\left[ \frac{\alpha_s(\mu)}{\alpha_s(\mu_0)} \right]^{-\gamma_0^m/(2 \beta_0)}
\bigg[ 1 - 
\frac{\gamma_1^m \beta_0 - \beta_1 \gamma_0^m}{2 \beta_0^2} 
\frac{\alpha_s (\mu) - \alpha_s(\mu_0)}{4 \pi} + \cdots \bigg] , 
\\
\frac{f_V^\perp (\mu) }{f_V^\perp (\mu_0)} 
&=& 
\left[ \frac{\alpha_s(\mu)}{\alpha_s(\mu_0)} \right]^{+\gamma_0^T/(2 \beta_0)}
\bigg[ 1 + 
\frac{\gamma_1^T \beta_0 - \beta_1 \gamma_0^T}{2 \beta_0^2} 
\frac{\alpha_s (\mu) - \alpha_s(\mu_0)}{4 \pi} + \cdots \bigg] , 
\end{eqnarray}
\end{subequations}
where 
\begin{subequations}
\begin{eqnarray}
&& 
\gamma_0^m = -6 C_F, \quad 
\gamma_1^m = -3 C_F^2 - \frac{97}{3} C_F C_A + \frac{20}{3} C_F T_F n_f, 
\\
&& 
\gamma_0^T = 2 C_F, \quad 
\gamma_1^T = -19 C_F^2 + \frac{257}{9} C_F C_A - \frac{52}{9} C_F T_F 
n_f.
\end{eqnarray}
\end{subequations}
Here, $\beta_0 = \frac{11}{3} N_c - \frac{2}{3} n_f$ is the one-loop
coefficient of the QCD beta function, $\beta_1 = \frac{34}{3} C_A^2 -
\frac{20}{3} C_A T_F n_f-4 C_F T_F n_f$ is the two-loop coefficient of
the QCD beta function, $C_F=(N_c^2-1)/(2N_c)$, $C_A=3$, $N_c=3$ is the
number of colors, $T_F=1/2$, and $n_f$ is the number of active quark
flavors.

\section{Evolution matrix \label{sec:evolution-matrix}}
At NLL accuracy, the evolution matrix $ U_{nk}(\mu,\mu_0)$ is given by 
\cite{Mueller:1993hg}
\begin{equation}
U_{nk}(\mu,\mu_0)=\left\{\begin{array}{ll}
E_n^{\rm NLO}(\mu,\mu_0),&\hbox{~if $k=n$,}\\
\frac{\textstyle\alpha_s(\mu)}{\textstyle 4\pi}
E_n^{\rm LO}(\mu,\mu_0)d_{nk}(\mu,\mu_0),&
\hbox{~if $k<n$,}
\end{array}
\right.
\label{Unk}
\end{equation}
where
\begin{subequations}%
\begin{eqnarray}
E_n^{\rm LO}(\mu, \mu_0)&=& 
\bigg[ \frac{\alpha_s (\mu)}{\alpha_s(\mu_0)}
\bigg]^{\frac{\gamma_n^{\perp(0)}}{2 \beta_0}},\\
E_n^{\rm NLO}(\mu, \mu_0) &=& E_n^{\rm LO}(\mu, \mu_0)
\bigg[ 1+ \frac{\alpha_s(\mu)-\alpha_s(\mu_0)}{4 \pi}
\frac{ \gamma_n^{\perp(1)} \beta_0 - \gamma_n^{\perp(0)} 
\beta_1 }{2 \beta_0^2} 
\bigg],\\
d_{nk}(\mu, \mu_0) &=& 
\frac{M_{nk}}{\gamma_n^{\perp(0)} - \gamma_k^{\perp(0)}-2 \beta_0}
\bigg\{ 
1- \bigg[ \frac{\alpha_s (\mu)}{\alpha_s(\mu_0)} 
\bigg]^{\frac{\gamma_n^{\perp(0)}-\gamma_k^{\perp(0)}-
2 \beta_0}{2 \beta_0}} \bigg\}, 
\\
M_{nk}
 &=& \frac{(k+1) (k+2) (2 n+3)}{(n+1) (n+2)} (\gamma_n^{\perp(0)} -
\gamma_k^{\perp(0)})
\nonumber \\ && 
\times 
\bigg[ 
\frac{8 C_F A_{nk}-\gamma_k^{\perp(0)}-2 \beta_0}{(n-k) (n+k+3)}
+4 C_F \frac{A_{nk}-\psi(n+2)+\psi(1)}{(k+1)(k+2)} \bigg],
\\ 
A_{nk} &=&
 \psi(\tfrac{n+k+4}{2})-\psi(\tfrac{n-k}{2})+2 \psi(n-k)
-\psi(n+2)-\psi(1).
\end{eqnarray}
\end{subequations}
Here $\psi(n)$ is the digamma function.
The LO and NLO anomalous dimensions,  
$\gamma_n^{\perp(0)}$ and $\gamma_n^{\perp(1)}$, respectively, are given by
\begin{subequations}%
\begin{eqnarray}
\gamma_n^{\perp(0)}&=&\gamma_n^{(0)}-\gamma_0^T,\\
\gamma_n^{\perp(1)}&=&\gamma_n^{(1)}-\gamma_1^T, 
\end{eqnarray}
\end{subequations}%
where, from Refs.~\cite{Lepage:1979zb,Shifman:1980dk}, we have
\begin{equation}
\gamma_n^{(0)}=8C_F(H_{n+1}-3/4),
\end{equation}
and, from Refs.~\cite{Vogelsang:1997ak,Hayashigaki:1997dn}, we have
\begin{subequations}
\begin{eqnarray}
\gamma_n^{(1)}
&\equiv&
4C_F^2
\left[
H^{(2)}_{n+1}-2H_{n+1}-\frac{1}{4}
\right]
+C_F C_A
\left[
-16H_{n+1}H^{(2)}_{n+1}
-\frac{58}{3}H_{n+1}^{(2)}
+\frac{572}{9}H_{n+1}
-\frac{20}{3}
\right]
\nonumber \\
&&
-8\left(C_F^2-\frac{1}{2}C_FC_A\right)
\bigg[
4H_{n+1}
\left(
S_{(n+1)/2}^{'(2)}-H_{n+1}^{(2)}-\frac{1}{4}
\right)
-8\tilde{S}_{n+1}
+S_{(n+1)/2}^{'(3)}
-\frac{5}{2}H_{n+1}^{(2)}
\nonumber \\
&&
+\frac{1+(-1)^n}{(n+1)(n+2)}+\frac{1}{4}
\bigg]
+\frac{32}{9}C_F \frac{n_f}{2}
\left[
3H_{n+1}^{(2)}-5H_{n+1}+\frac{3}{8}
\right],
\end{eqnarray}
where 
\begin{eqnarray}
H_{n}^{(k)}&\equiv&\sum_{j=1}^n \frac{1}{j^k},
\quad
\textrm{with}
\quad
H_n^{(1)}\equiv H_{n},
\\
S^{'(k)}_{n/2}&\equiv&
\begin{cases}
\displaystyle
H^{(k)}_{n/2},
& \textrm{if $n$ is even,}
\\
\displaystyle
H^{(k)}_{(n-1)/2},
& \textrm{if $n$ is odd,}
\end{cases}
\\
\tilde{S}_n
&\equiv&
\sum_{j=1}^n \frac{(-1)^j}{j^2}H_j.
\end{eqnarray}
\end{subequations}
Here, the $H_n^{(k)}$ are the generalized harmonic numbers.
Note that the off-diagonal matrix elements, which are proportional to
$d_{nk} (\mu, \mu_0)$, are nonvanishing only for even
$n-k$~\cite{Mueller:1993hg, Mueller:1994cn}. One can obtain
$U_{nk}(\mu,\mu_0)$ at LL accuracy by replacing $E_n^{\rm
NLO}(\mu,\mu_0)$ in Eq.~(\ref{Unk}) with $E_n^{\rm LO}(\mu,\mu_0)$ and
setting the off-diagonal terms to zero.

\section{Nonrelativistic expansion\label{sec:nonrel-expansion}}

In this appendix we discuss the nonrelativistic expansion of the
light-cone amplitude in order $\alpha_s^0$ and investigate the
convergence of that expansion numerically.

\subsection{Formulation of the expansion\label{sec:form-nonrel-expansion}}

In Ref.~\cite{Bodwin:2014bpa}, a formal expansion of the LCDA was 
given. Making the change of light-cone variables $x\to 2x-1$, we write 
that expansion as 
\begin{equation}
\phi_V^\perp(x)=
\sum_{k=0}^\infty \frac{(-1)^k\langle x^{k}\rangle}{2^{k}k!}
\delta^{(k)}(x-\tfrac{1}{2}),
\label{nonrel-expansion}
\end{equation}
where the normalization condition is
\begin{equation}
\int_0^1 dx\, \phi_V^\perp(x)=1.
\end{equation}
Here,
$\langle x^{k}\rangle$ is defined by
\begin{equation}
\label{eq:x-mom-def}
\langle x^{k}\rangle = 2^{k}\int_0^1 dx\,
(x-\tfrac{1}{2})^{k} \phi_V^\perp(x).
\end{equation}
As we will see in Appendix~\ref{sec:nonrel-mom-sizes}, the $k$th $x$
moment in Eq.~(\ref{eq:x-mom-def}) is proportional, in order
$\alpha_s^0$, to the NRQCD LDME $\langle v^k\rangle$. Hence, the
expansion in Eq.~(\ref{nonrel-expansion}) is the nonrelativistic
expansion of the LCDA in order $\alpha_s^0$. In the following
discussions, we will assume that $\phi_V^\perp(x)$ is even under the
replacement $x\leftrightarrow 1-x$ (charge-conjugation parity), in which
case, only the moments $\langle x^k\rangle$ with $k$ even are
nonvanishing.

The meaning of this formal expansion is that, if one integrates 
$\phi_V^\perp(x)$ against a test function $f(x)$, then that integral is 
replaced by the sum  of the integrals of $\phi_V^\perp(x)$ against each 
term in the Taylor expansion of $f(x)$,
\begin{subequations}%
\begin{equation}
\int_0^1 dx\, f(x)\phi_V^\perp(x) 
=\sum_{k=0}^\infty \frac{1}{k!} 
\biggl[\frac{d^k}{dx^k}f(x)\biggr]\biggr|_{x=1/2}
\int_0^1 dx\, (x-1/2)^k \phi_V^\perp(x)
=\sum_{k=0}^\infty f^{(k)} \langle x^k\rangle,
\label{Taylor-expansion}
\end{equation}
where
\begin{equation}
f^{(k)}=\frac{1}{2^k k!} \biggl[\frac{d^k}{dx^k}f(x)\biggr]\biggr|_{x=1/2}.
\label{f-k}
\end{equation}
\end{subequations}%

In our case, we wish to compute the light-cone amplitude
\begin{equation}
{\cal M}^{(0)}(\mu)=\int_0^1 dx\, T_H(x,\mu)\phi_{V}^\perp(x,\mu),
\label{integral-amplitude}
\end{equation}
where the superscript $(0)$ denotes order $\alpha_s^0$. ${\cal 
M}^{(0)}(\mu)$ has the nonrelativistic expansion
\begin{equation}
{\cal M}^{(0)}(\mu)=\sum_{k=0}^{\infty} {\cal M}^{(0,v^{2k})}(\mu),
\label{nonrel-expansion-amplitude}
\end{equation}
where
\begin{equation}
{\cal M}^{(0,v^{2k})}(\mu)=f^{(2k)} \langle x^{2k}\rangle,
\end{equation}
and we make the identification
\begin{equation}
f(x)=\sum_{n=0}^\infty\sum_{m=0}^n 
T_n(\mu) U_{nm}(\mu,\mu_0) N_m C_m^{(3/2)}(2x-1).
\label{Gegenbauer-TH}
\end{equation}
We compute the derivatives of this quantity by making use of the Abel
summation in Eq.~(\ref{limit-eigenvalue-sum}). That is, we compute
\begin{equation}
f^{(2k)}
=
\lim_{z\to1}
\sum_{n=0}^\infty 
\sum_{m=0}^n z^n
T_n(\mu) U_{nm}(\mu,\mu_0) 
N_m 
\frac{1}{2^{2k}(2k)!} \frac{d^{2k}}{dx^{2k}}
C_m^{(3/2)}(2x-1)\bigg|_{x=1/2}.
\label{nonrel-coeffs}
\end{equation}
and we accelerate the convergence of the sum of $m$ by making use of
Pad\'e approximants, as we have described earlier.

Making use of the identities
\begin{subequations}%
\begin{equation}
\frac{d}{dx}C_n^{\lambda/2}(x)
=
\lambda C_{n-1}^{(\lambda+2)/2}(x)
\end{equation}
and 
\begin{equation}
C_{2n}^{\lambda/2}(0)
=
\frac{(-1)^n}{(2n)!!}
\frac{(\lambda+2n-2)!!}{(\lambda-2)!!},
\end{equation}
\end{subequations}%
we obtain a convenient expression for the even derivatives of the
even-order Gegenbauer polynomials:
\begin{equation}
\frac{d^{2k}}{dx^{2k}} C_{2n}^{(3/2)}(2x-1)\biggr|_{x=1/2}
=(-1)^{n-k}2^{2k}\frac{(2n+2k+1)!!}{(2n-2k)!!}.
\end{equation}

\subsection{Sizes of the nonrelativistic moments\label{sec:nonrel-mom-sizes}}

In order $\alpha_s^0$, the $x$ moments of the LCDA
[Eq.~(\ref{eq:x-mom-def})]  have the following relationships to the NRQCD
LDMEs \cite{Braguta:2007fja,Braguta:2007fh,Braguta:2012zza}:
\begin{equation}
\langle x^{2k}\rangle =\frac{\langle v^{2k}\rangle}{2k+1}.
\label{mom-LDME}
\end{equation}
As we have mentioned in footnote~\ref{footnote:v-scaling}, the NRQCD
velocity-scaling rules, in their strictest sense, state that $\langle
v^n\rangle_V$ vanishes as $v^n$ in the limit $v\to 0$. However, in
phenomenology, the velocity-scaling rules are usually taken to mean
that
\begin{equation}
\langle v^{2k}\rangle\sim \langle v^2\rangle^k,
\label{velocity-scaling}
\end{equation}
where $\sim$ means equal up to a coefficient of order 1.
These approximate sizes of the LDMEs are consistent with the
generalized Gremm-Kapustin relation \cite{Bodwin:2006dn}.

Now let us consider the $x$ moments of the model LCDA in
Eq.~(\ref{KN-LCDA}), which we denote by $\langle x^{k}\rangle_M$.
We compute $\sigma_{J/\psi}(\mu_0)$ using Eq.~(\ref{KNLDME-2dmom}), but
we drop the order-$\alpha_s$ term so as to obtain the behavior at order
$\alpha_s^0$. Then, using $\langle v^2\rangle_{J/\psi}=0.201$ we obtain
$\sigma_{J/\psi}=0.129422$. The first several $x$ moments are then
\begin{eqnarray}
\label{eq:KN-xi-2k}
\langle x^{0}\rangle_M
&=&1,
\nonumber \\
\langle x^{2}\rangle_M
&=&0.0573955,
\nonumber \\
\langle x^{4}\rangle_M
&=&0.00962303,
\nonumber \\
\langle x^{6}\rangle_M
&=&0.00259973,
\nonumber \\
\langle x^{8}\rangle_M
&=&0.000943655,
\nonumber \\
\langle x^{10}\rangle_M
&=&0.000419855.
\end{eqnarray}
On the other hand, from the relationship between the $x$
moments and the LDMEs at order $\alpha_s^0$ [Eq.~(\ref{mom-LDME})] and
the NRQCD velocity-scaling rules [Eq.~(\ref{velocity-scaling})], we
expect that
\begin{eqnarray}
\langle x^{0}\rangle 
&=&1,
\nonumber\\
\langle x^{2}\rangle
&=&0.0573955,
\nonumber \\
\langle x^{4}\rangle
&\sim&0.00592964,
\nonumber \\
\langle x^{6}\rangle
&\sim&0.000729288,
\nonumber \\
\langle x^{8}\rangle
&\sim&0.0000976684,
\nonumber \\
\langle x^{10}\rangle
&\sim&0.0000137595.\label{v-scaling-moments}
\end{eqnarray}
The expression for $\langle x^{2k}\rangle_M$ in the limit $\sigma_V\to
0$ is given by
\begin{equation}
\langle 
x^{2k}\rangle_M=(2\sigma_V)^{2k}(2k-1)!!\,\bigl[1+O(\sigma_{V}^2)\bigr]. 
\label{sigma-to-0}
\end{equation}
Hence, the model LCDA satisfies the NRQCD velocity-scaling rules in the
strict sense that the $2k$th moment vanishes as the $k$th power of a
quantity that could be interpreted as the square of the velocity.
However, we see from Eq.~(\ref{eq:KN-xi-2k}) that the first several $x$
moments of the model LCDA badly violate the broader expectation that the
LDMEs satisfy the relationship in Eq.~(\ref{v-scaling-moments}).

The crucial issue for the convergence of the velocity expansion is the
behavior of the $2k$th $x$ moment of the LCDA in the limit $k\to \infty$
for fixed $\sigma_V$. We can derive an asymptotic expansion for the
$x$ moments of the model LCDA by integrating the definition in
Eq.~(\ref{eq:x-mom-def}) twice by parts. The result for even moments is
\begin{eqnarray}
\langle x^{2k}\rangle_M&=&\frac{\bigl[\frac{\partial}{\partial x}\phi_V^{\perp 
M}(x)\bigr]\bigr|_{x=0}-\bigl[\frac{\partial}{\partial x}\phi_V^{\perp 
M}(x)\bigr]\bigr|_{x=1}}{4(2k+1)(2k+2)}+O[1/(2k)^3]\nonumber\\
&=&N_\sigma \frac{\sqrt{2/\pi}e^{-1/(8\sigma_V^2)}}
{\sigma_V(2k+1)(2k+2)}+O[1/(2k)^3].
\label{k-to-infinity}
\end{eqnarray}
Hence, we see that the $2k$th moment falls as $1/k^2$ in the limit $k\to
\infty$, while we expect, from Eqs.~(\ref{mom-LDME}) and
(\ref{v-scaling-moments}), that the $2k$th moment should fall faster
than $v^{2k}$. Nevertheless, Eq.~(\ref{k-to-infinity}) shows that the
nonrelativistic expansion converges for the model LCDA, in the 
absence of evolution, provided that
\begin{equation}
T_H^{(2k)}\equiv \frac{1}{2^{2k} (2k)!} 
\biggl[\frac{d^{2k}}{dx^{2k}}T_H(x)\biggr]\biggr|_{x=1/2} 
\end{equation}
grows more slowly than a power of $k$.\footnote{We note that the
limits $k\to\infty$ and $\sigma_V\to 0$ cannot be interchanged, as can
be seen explicitly from Eqs.~(\ref{sigma-to-0}) and
(\ref{k-to-infinity}).}

We record here the values for $\langle x^{2k}\rangle$ that we obtain by 
retaining both the order-$\alpha_s$ term and the order-$v^2$ term in 
Eq.~(\ref{KNLDME-2dmom}), which corresponds to taking 
$\sigma_{J/\psi}=0.228$. 
\begin{eqnarray}
\label{eq:KN-sigma-0228}
\langle x^{0}\rangle_M
&=&1,
\nonumber \\
\langle x^{2}\rangle_M
&=&0.120256,
\nonumber \\
\langle x^{4}\rangle_M
&=&0.0373473,
\nonumber \\
\langle x^{6}\rangle_M
&=&0.0166916,
\nonumber \\
\langle x^{8}\rangle_M
&=&0.00909954,
\nonumber \\
\langle x^{10}\rangle_M
&=&0.00561735.
\end{eqnarray}
These $x$ moments, of course, lead to a slower convergence 
of the nonrelativistic expansion than those for the case 
$\sigma_{J/\psi}=0.129422$.

\subsection{Numerical tests of the convergence of the nonrelativistic 
expansion\label{sec:nonrel-num-tests}}

Now let us test numerically the convergence of the nonrelativistic
expansion of the light-cone amplitude in order $\alpha_s^0$, which
is given in Eq.~(\ref{nonrel-expansion-amplitude}). We do this by
comparing the numerical results from the nonrelativistic expansion of
the light-cone amplitude with the numerical results that are obtained by
computing the light-cone amplitude directly from a model LCDA. For this
purpose, we make use of the model LCDA in Eq.~(\ref{KN-LCDA}). As we
have pointed out, the $x$ moments of this model LCDA decrease much more
slowly with increasing moment number than would be expected from the
NRQCD velocity-scaling rules. Therefore, we expect the nonrelativistic
expansion to converge more slowly for this model LCDA than for a more
realistic LCDA. However, as we will see, even for this model LCDA, the
convergence of the nonrelativistic expansion is quite rapid.

\subsubsection{Without evolution}

We first take the case of no evolution, {\it i.e.}, $\mu=\mu_0$. We 
consider $T_H(\mu)$ at leading order in $\alpha_s$. Then, 
$f(x)=T_H^{(0)}$, and we can compute $f^{(2k)}$ analytically from 
Eq.~(\ref{f-k}), with the result 
\begin{equation}
\mathcal{M}^{(0,v^{2k})}(\mu_0)= 4\langle x^{2k}\rangle
\label{M-LO-no-evolve}
\end{equation}
for all $k$. We note that we can also compute the $f^{(2k)}$ in
Eq.~(\ref{M-LO-no-evolve}) by making use of the Abel summation in
Eq.~(\ref{nonrel-coeffs}). If we accelerate the convergence of the sum
over $m$ by employing a $50\times 50$ Pad\'e approximant, then, through
${\cal M}^{(0,v^{10})}$, the agreement with the coefficient $4$ in
Eq.~(\ref{M-LO-no-evolve}) holds to greater than $5$ places after
the decimal. This agreement provides strong confirmation of the 
validity of the Abel summation in Eq.~(\ref{nonrel-coeffs}), as 
supplemented by the use of Pad\'e approximants.
 
Using the $x$ moments of the model LCDA that correspond to
$\sigma_{J/\psi}=0.129422$ [Eq.~(\ref{eq:KN-xi-2k})], we find that
\begin{equation}
\sum_{k=0}^5 \mathcal{M}^{(0,v^{2k})}(\mu_0)|_{M}=4.28393.
\end{equation}
On the other hand, if we evaluate $\mathcal{M}(\mu_0)$ directly in
Gegenbauer-moment space, taking the first 20 Gegenbauer moments, we
obtain
\begin{equation}
\mathcal{M}(\mu_0)|_{M}
\approx 4.28670.
\end{equation}
This value agrees very well with the one that is obtained from the first
5 terms in the nonrelativistic expansion. [It also agrees very well with
the value that is obtained by direct computation of the amplitude in $x$
space as, in Eq.~(\ref{integral-amplitude}).] The order-$v^2$ term in
the expansion accounts for $80\%$ of the higher-order corrections. As we
have noted, the $x$ moments of the model LCDA severely violate the
velocity-scaling relation in Eq.~(\ref{velocity-scaling}), and, so,
we would expect that, in the case of a more realistic LCDA, the
order-$v^2$ term in the expansion would account more fully for the
higher-order corrections. If we use the values of the $x$ moments in
Eq.~(\ref{v-scaling-moments}), which are based on the NRQCD
velocity-scaling rules, then we find that the order-$v^2$ term in the
expansion accounts for $89\%$ of the higher-order corrections.

We can evaluate these same quantities for the $x$ moments in
Eq.~(\ref{eq:KN-sigma-0228}), which correspond to the choice
$\sigma_{J/\psi}=0.228$. We remind the reader that this value of
$\sigma_{J/\psi}$ corresponds to the inclusion of the order-$\alpha_s$
corrections, as well as the order-$v^2$ corrections, in the model LCDA.
Hence, for this value of $\sigma_{J/\psi}$, the relationship between the
$x$ moments of the model LCDA and the NRQCD LDMEs in
Eq.~(\ref{mom-LDME}) does not hold, and the $x$-moment expansion is not,
strictly speaking, a nonrelativistic expansion. Nevertheless, it is
interesting to examine the convergence of the $x$-moment expansion in
this case. The result for the $x$-moment expansion is
\begin{equation}
\sum_{k=0}^5 \mathcal{M}^{(0,v^{2k})}(\mu_0)|_{M}=4.75605,
\end{equation}
and the result for the direct evaluation, using the first 20 Gegenbauer
moments, is
\begin{equation}
\mathcal{M}(\mu_0)|_{M}
\approx
4.84334.
\end{equation}
Again, there is good agreement between the results from $x$-moment
expansion and the direct evaluation, although, as expected, the
$x$-moment expansion converges more slowly with this choice of
$\sigma_{J/\psi}$. 

\subsubsection{With evolution}

In this section, we compute the same quantities as in the preceding
section, but taking $\mu=m_H$ and $\mu_0=1$~GeV. We use LL evolution. In
order to compute the coefficients of the $\langle x^{2k}\rangle$ in
the presence of evolution, we use the Abel summation in
Eq.~(\ref{nonrel-coeffs}), accelerating the convergence to the limit by
employing a $50\times 50$ Pad\'e approximant. The result is
\begin{eqnarray}
\label{eq:KN-M-sigma-no-alphas}
\mathcal{M}^{(0,0)}(\mu)
&=&
4.91403\langle x^{0}\rangle,
\nonumber \\
\mathcal{M}^{(0,v^2)}(\mu)
&=&
2.95670\langle x^{2}\rangle,
\nonumber \\
\mathcal{M}^{(0,v^4)}(\mu)
&=&
2.31150\langle x^{4}\rangle,
\nonumber \\
\mathcal{M}^{(0,v^6)}(\mu)
&=&
1.96596\langle x^{6}\rangle,
\nonumber \\
\mathcal{M}^{(0,v^8)}(\mu)
&=&
1.74271\langle x^{8}\rangle,
\nonumber \\
\mathcal{M}^{(0,v^{10})}(\mu)
&=&
1.58320\langle x^{10}\rangle.
\end{eqnarray}
We note that the evolution results in a decreasing sequence of 
coefficients, and, so we expect the nonrelativistic expansion to 
converge more rapidly than in the absence of evolution. 
With choice $\sigma_{J/\psi}=0.129422$, the nonrelativistic
expansion gives
\begin{equation}
\sum_{k=0}^5 \mathcal{M}^{(0,v^{2k})}(\mu)|_M=5.11340,
\end{equation}
and the direct evaluation, using the first 20 Gegenbauer
moments, gives 
\begin{equation}
\mathcal{M}(\mu)|_{M}
=5.11425.
\end{equation}
There is good agreement between the nonrelativistic expansion and the
direct evaluation. As expected, the nonrelativistic expansion converges
more rapidly than in the case of no evolution. In this case, the
order-$v^2$ term in the expansion accounts for $85\%$ of the higher-order
corrections. We would expect that, in the case of a more realistic LCDA,
the order-$v^2$ term in the expansion would account more fully for the
higher-order corrections. If we use the values of the $x$ moments in
Eq.~(\ref{v-scaling-moments}), which are based on the NRQCD
velocity-scaling rules, then we find that order-$v^2$ term in the
expansion accounts for $92\%$ of the higher-order corrections.

Finally, we carry out the same computation with the choice
$\sigma_{J/\psi}=0.228$. Again, we remind the reader that this value
of $\sigma_{J/\psi}$ corresponds to the inclusion the order-$\alpha_s$
corrections, as well as the order-$v^2$ corrections, in the model LCDA,
and, so, for this value of $\sigma_{J/\psi}$, the expansion the
$x$-moment expansion of the LCDA is not, strictly speaking, a
nonrelativistic expansion. The result for the $x$-moment expansion is
\begin{equation}
\sum_{k=0}^5 \mathcal{M}^{(0,v^{2k})}(\mu)|_M=5.41349,
\end{equation}
and the result from the direct evaluation is 
\begin{equation}
\mathcal{M}(\mu)|_{M}
\approx
5.43700.
\end{equation}
Again, the $x$-moment expansion converges rapidly to the result
from the direct evaluation, although, as expected, not as rapidly as
with the choice $\sigma_{J/\psi}=0.129422$.

\begin{acknowledgments}                                      

We thank Deshan Yang for clarifying several issues with regard to the
formulas in Ref.~\cite{Wang:2013ywc}. The work of G.T.B.\ and H.S.C.\
is supported by the U.S.\ Department of Energy, Division of High Energy
Physics, under Contract No. DE-AC02-06CH11357. The work of J.-H.E.\ is
supported by Global Ph.D. Fellowship Program through the National
Research Foundation (NRF) of Korea funded by the Korean government (MOE)
under Grant No.\ NRF-2012H1A2A1003138. J.L.\ thanks the Korean Future
Collider Working Group for enjoyable discussions regarding the work
presented here. The work of J.L.\ was supported by the Do-Yak project of
NRF under Contract No.\ NRF-2015R1A2A1A15054533. The submitted
manuscript has been created in part by UChicago Argonne, LLC, Operator
of Argonne National Laboratory. Argonne, a U.S.\ Department of Energy
Office of Science laboratory, is operated under Contract No.
DE-AC02-06CH11357. The U.S. Government retains for itself, and others
acting on its behalf, a paid-up nonexclusive, irrevocable worldwide
license in said article to reproduce, prepare derivative works,
distribute copies to the public, and perform publicly and display
publicly, by or on behalf of the Government.

\end{acknowledgments}


\end{document}